\documentclass[12pt,a4paper,english,twoside,onecolumn]{iopart}
\pdfoutput=1
\usepackage{graphicx}
\usepackage{setstack}
\usepackage{bm}
\usepackage{color}

\newcommand{\laeq}{\raisebox{-.7ex}{$\stackrel{\textstyle<}{\sim}$}\ }

\begin{document}
%

\title[From the orbit theory to a GC parametric equilibrium distribution function]{From the orbit theory to a guiding center parametric equilibrium distribution function}
\author{C. Di Troia}

\address{Associazione Euratom-ENEA sulla Fusione, CR Frascati, CP 65 - 00044 Frascati, Italy}
 \ead{claudio.ditroia@enea.it}

 
 \begin{abstract}
 This work proposes a parametric equilibrium distribution function $\mathcal{F}_{eq}$  to be applied to the gyrokinetic studies of the Finite Orbit Width behavior of guiding centers representing several species encountered in axisymmetric tokamak plasma, as  fusion products, thermal bulk and energetic particles from Ion Cyclotron Radiation Heating and Negative Neutral Beam Injections.

After the analysis of the basic results of orbit theory obtained with a particularly convenient orbit coordinates set, it is shown how the proposed $\mathcal{F}_{eq}$ satisfies the two conditions that make it an equilibrium distribution function: (i) it must depend only on the constants of motion and adiabatic invariants, and (ii) the guiding centers must remain confined for suitably long time. 

Furthermore, the $\mathcal{F}_{eq}$ can be modeled, with a proper choice of its  parameters, to reproduce the most common distribution functions. A local Maxwellian distribution function is obtained for the thermal plasma in the Zero Orbit Width approximation.
 For the fusion $\alpha$ particles, $\mathcal{F}_{eq}$ can also reproduce the Slowing Down (SD) distribution function. More generally, for supra-thermal particles, when external heatings  are present, such as (N)NBI and ICRH,  the proposed model distribution function shows similarities with the anisotropic SD  and the biMaxwellian distribution functions.

$\mathcal{F}_{eq}$ can be used to fit experimental profiles and it could provide a useful tool for experimental and numerical data analysis. Moreover, it could help to develop analytical computations for facilitating data interpretation in the light of theoretical models. This distribution function can be easily implemented in  gyrokinetic codes, where it can be used to simulate plasma also in the presence of external heating sources. 

\end{abstract}



\section{Introduction}
\label{sec:intro}
Gyrokinetic theory in general and gyrokinetic simulations in particular make often use of an initial distribution function of guiding centers (GCs), usually indicated by $F_0$. Initial distribution function must represent a slowly evolving equilibrium for a perturbative approach. There are two conditions that $F_0$ must satisfy in order to be considered an equilibrium $\mathcal{F}_{eq}$ one:
\begin{equation}
\label{1}
\mbox{it must depend only on the invariants of motion}
\end{equation}
and
\begin{equation} 
\label{2}
\mbox{the GCs must remain confined for suitably long time.}
\end{equation}
 The meaning of the first sentence is clear: the function of constants is constant. The second condition takes into account  that laboratory plasma is placed into a limited portion of space; if the particles are
contained within this region only for a finite time interval and then lost, consequently the corresponding particles distribution will be impoverished in the protracted time.

The work is set up to progressively build  the equilibrium distribution function. In the next section, it is shown the chosen set of constants of motion (COM) in order to describe the unperturbed orbits characterizing the equilibrium. There is no need of entering into the details of the Guiding Center (GC) transformation. The reader who is not confident with this coordinate transformation can consult \cite{deblank} or one of the many introductory articles on this topic. Neverthless, it is explained the physical assumptions needed to consider as constants the following GC quantities: $\mathcal{P}_\phi$ (corresponding to the axisymmetry angular momentum), the kinetic energy per unit mass $w$ and the generalized pitch angle variable $\lambda$, defined as the ratio between the magnetic moment $\mu$ and $w$. In the text, the set ($\mathcal{P}_\phi,w,\lambda$) will be referred to as the Quasi Invariants (QIs) set.  The Section (\ref{subsec:problems}) provides a brief outline on how it is currently solved the problem of assigning an equilibrium distribution function in gyrokinetic codes \cite{idomura,fu,angelino,garbet,grangirard,dif,vernay}. The difficulties arising from that common procedure will be the starting point for the developing of an alternative approach.
Section \ref{sec:orb} addresses  the above condition (\ref{2}), where a GC is considered lost when a portion of its unperturbed orbit will be out of the plasma section. The main topic of this Section is devoted to specify the topology of the projected orbits onto the poloidal section. Indeed, to better understand an equilibrium distribution function of QIs, it is convenient to consider it as a distribution of (projected) orbits. It will be useful to know if GCs are, not only confined or lost, but also, \emph{trapped} or \emph{passing}.

In this analysis, the poloidal magnetic flux coordinate $\psi$, in addition to $\mathcal{P}_\phi, w$ and $\lambda$, are used as orbit coordinates. In the light of this set of coordinates, it will be convenient to show the basic results of the orbit theory \cite{rome,putv,hsu,egedal,eriksson01,white,chiu}, such as the orbits classification, the derivation of a graphical method to promptly deduce the orbit shape, the definition of the orbit average, the expression of the characteristic frequencies (\emph{transit} and \emph{bounce} frequencies) and the expression of the \emph{second invariant}.

Finally, in Section \ref{sec:good} the parametric $\mathcal{F}_{eq}$ equilibrium distribution function is built adopting the following guidelines: it must have Boltzmann-like behavior, the energy must be represented in terms of QIs and it must be mathematically tractable. Here, the similarities found in various limiting behaviors are represented: when the proposed $\mathcal{F}_{eq}$ is compared with the commonly used distribution functions in tokamak plasma physic such as the local Maxwellian, the Slowing Down (SD), the biMaxwellian distribution function and so on.

Section \ref{sec:qual} qualitatively describes the various terms that constitute the obtained $\mathcal{F}_{eq}$. 

An appendix has been inserted solely to support the reader in order to easily refer to some known formulas for the description of the magnetic field with the adopted convention.  Appendix A briefly considers two particularly used coordinates reference systems: \emph{magnetic flux coordinates} and \emph{Shafranov coordinates}. The Shafranov geometry, characterized by circular plasma poloidal section, is the example geometry adopted to visualize some results useful for more general plasma poloidal sections. 

In this work the vectors are not indicated with bold letters. As for example, the guiding center velocity $V$, the electric field $E$, the magnetic field $B$, the unit vector $b$ in the direction of the magnetic field, the perpendicular (to the magnetic field) velocity $v_\perp$, the drift velocity $v_D$, and the \emph{grad} operator $\nabla$ are vector quantities, whilst the parallel (to the magnetic field) velocity $v_\| \in [-\infty,+\infty]$, the velocity magnitude $v$ and the radial coordinate $r$ are scalars. The symmetry axis component of the angular momentum $L_Z$ and the toroidal component of the canonical angular momentum $\mathcal{P}_\phi$ are vector components. 
When there is a possibility of confusion it will be specified the scalar or vector character of the adopted symbol. 

Moreover, \emph{natural units} (n.u.) are employed: the speed of light $c=1$.

\section{Preliminary concepts}
\label{sec:concepts}
It is well known that the particle energy $\mathcal{E}= m_s v^2/2+q_s A_0$ (where $A_0$ is the electric potential and $m_s, q_s$ specify the considered particle species with its mass and its charge) and the (canonical) angular momentum component along the symmetry axis $L_Z=m_s R v_\phi + q_s R A_\phi$ (where $A_\phi$ and $v_\phi$ are respectively the toroidal component of the magnetic vector potential $A$ and of the particle velocity $v$) are COMs, assuming a charged particle ($m_s,q_s$) in non relativistic regime,  the toroidal symmetry and the presence of the only Electro-Magnetic field.
$L_Z$ is frequently named \emph{canonical toroidal momentum}.  Finally the magnetic moment $\mu=v_\perp^2 /(2 |B|)$ is an adiabatic invariant.

These particle invariants undergo the Guiding Center (GC) transformation\footnote[1]{The reason why it is used the GC instead of the gyrocenter transformation is because it is analyzed the equilibrium, where the fields are unperturbed.} which reduces the dimensionality of the velocity space from 3D to 2D, removing the gyroangle $\gamma$, the angle that fixes the direction of $v_\perp$, up to a wanted order in $\delta=\rho/L$ (where $\rho$ is the Larmor radius and $L$ is a typical scale length of the system). As for the gyroangle, also the conservation property of the transformed GC COM is usually verified up to a given order in $\delta$. Thus, preserving the same symbols and fixing a given $\delta$ order, $L_Z, \mathcal{E}$ and $\mu$ are (approximate) GC COMs. 

Here a slightly different viewpoint is adopted: the constants of motion in GC coordinates are exact constants, whilst the corresponding quantities, expressed in particle coordinates, become  approximate.
In a schematic manner, it is common practice considering
$$
\mbox{(exact) COM in particle coord.=(approximate) COM in GC coord. + $O(\delta^n)$}
$$
while in this paper
$$
\mbox{(exact) COM in GC coord.=(approximate) COM in particle coord. + $O(\delta^n)$}.
$$
 The results must be correct up to the (n-1)-th $\delta$ order for both descriptions. Although the first and most used approach is more realistic, the second one is preferable when there is no need to come back to the particle coordinates once arrived at the GC description. Ref.\cite{brizard09} uses the same approach referring to COM in unperturbed GC dynamical reduction system. The use of the second approach is clearer than the first one because it fulfills the exact conservation of some GC quantities. The imposed constants of motion for the GC description are renamed QI to differentiate the two approaches: Quasi Invariants are exact Constants Of Motions in GC coordinates.
 
\subsection{Quasi Invariants}
\label{subsec:QI}
 The unperturbed GC projected orbit onto a poloidal section will be described in the toroidal coordinates $r$, $\theta$ and $\phi$, by the following relations: 
\begin{equation}
\label{eqmot}
\eqalign{
\dot{r} = V\cdot \nabla r,
\\ \dot{\theta} = V\cdot \nabla \theta, 
}
\end{equation} 
while the toroidal motion is described by:
\begin{equation}
\label{phimot}
\dot{\phi}=V\cdot \nabla \phi,
\end{equation}
 where $V$ is the GC velocity. Here, the definition of the poloidal angle $\theta$ is quite arbitrary: it is a $2\pi$ periodic coordinate and it lies in a plane orthogonal to the toroidal unit vector $e_\phi$. Even the definition of the radial coordinate can differ but it is required labeling the magnetic flux: $r=r(\psi)$.
 
 When the $E \times B$ drift, the $\nabla B$ drift and the curvature drift is retained correctly whilst the $O(\delta^2)$ drift is neglected, then $V$ can be expressed as:
 \begin{equation}
 \label{vGC}
V=\mathcal{Q} v_\| b+v_D=\mathcal{Q} \left[ v_\| b + \frac{m_sv_\|}{q_s|B|}\nabla \times (v_\|b) \right],
\end{equation}
where $b=B/|B|$,  $v_D$ is the drift velocity and $\mathcal{Q}=1+O(\delta)$ depends on how have been defined $v_\|$ in the GC transformation: commonly  $v_\|= V\cdot b$ and $\mathcal{Q}= 1+m_s/(q_s|B|)b\cdot \nabla \times (v_\|b)$ as in \cite{littleJ83,Balescu}, otherwise if $v_\| = (V-v_D)\cdot b$ then $\mathcal{Q}= 1$ recovering the original expression proposed by \cite{morozov}.

The equations in (\ref{eqmot}) are explicited as follows:
\begin{equation}
\label{a}
\eqalign{\dot{r}&=\mathcal{Q}\frac{m_sv_\|}{q_s |B|}\nabla \times (v_\|b)\cdot \nabla r
\\ \dot{\theta}&=\mathcal{Q} \left[v_\|b \cdot \nabla \theta+\frac{m_sv_\|}{q_s|B|}\nabla \times (v_\|b)\cdot \nabla \theta \right].
}
\end{equation}
For convenience, the magnetic field is written as follows (\ref{B0}): 
\begin{equation}
B=\nabla \psi \times \nabla \phi + F(\psi)\nabla \phi,
\end{equation}
where the radial component of the plasma current density is required to be zero to ensure force balance (otherwise $F(\psi)$ should be replaced by $F(\psi,\theta)$, see (\ref{Jr})). 

The terms  $b\cdot\nabla \theta, \nabla \times (v_\|b)\cdot \nabla \theta$ and $\nabla \times (v_\|b)\cdot \nabla r$, thanks to the $\phi$ symmetry, becomes:
\begin{equation}
\label{b}
b\cdot\nabla \theta=\frac{B}{|B|}\cdot \nabla \theta=\frac{\nabla\psi\times\nabla\phi\cdot\nabla\theta}{|B|}=-\frac{\psi^\prime}{|B|\sqrt{g}}
\end{equation}
\begin{equation}
\label{c}
\nabla \times (v_\|b)\cdot \nabla \theta=\nabla(\frac{v_\|F}{|B|})\times \nabla \phi\cdot \nabla \theta=-\frac{1}{\sqrt{g}}\partial_r(\frac{v_\|F}{|B|})
\end{equation}
\begin{equation}
\label{d}
\nabla \times (v_\|b)\cdot \nabla r=\nabla(\frac{v_\|F}{|B|})\times \nabla \phi\cdot \nabla r=\frac{F}{\sqrt{g}}\partial_\theta(\frac{v_\|}{|B|})
\end{equation}
where $\sqrt{g}=(\nabla r\times\nabla\theta\cdot\nabla\phi)^{-1}$ is the spatial Jacobian for the toroidal coordinates transformation. $\psi=\psi(r)$ is invertible\footnote{For simplicity, the magnetic flux surfaces are considered nested.} and the $prime$ indicates the radial derivative with $\psi^\prime<0$. 
Because $\nabla \theta$ together with $\nabla r$ is orthogonal to $\nabla \phi = e_\phi/R$, then $\nabla \times (v_\||B|^{-1}\nabla\psi\times\nabla\phi)$ is parallel to $e_\phi$ (it does not give any contribution if scalarly multiplied per $\nabla \theta$ or $\nabla r$). From (\ref{a},\ref{b},\ref{c},\ref{d}) and with $\dot{r}= \dot{\psi}/\psi^\prime$ and $\tilde{F}=m_sF/q_s$, the following relations are obtained:
\begin{equation}
\label{dotpsi}
\dot{\psi}=\psi^\prime \mathcal{Q}\frac{\tilde{F}v_\|}{|B|\sqrt{g}}\partial_\theta \frac{v_\|}{|B|}=-\frac{\psi^\prime \tilde{F}\partial_\theta(v_\|/|B|)}{\psi^\prime+\partial_r(\tilde{F}v_\|/|B|)} \dot{\theta}.
\end{equation}
The constancy of $\mathcal{P}_\phi$ defined as $\mathcal{P}_\phi=\psi+\tilde{F}v_\|/|B|$ is shown putting $\dot{\psi}$ on the RHS, changing the sign and multiplying for the denominator: 
\begin{equation}
0=\dot{\psi}+\dot{\psi} \partial_\psi \frac{\tilde{F}v_\|}{|B|}+\dot{\theta} \tilde{F} \partial_\theta \frac{v_\|}{|B|}=\dot{\psi}+\frac{d}{dt}\frac{\tilde{F}v_\|}{|B|}=\frac{d}{dt}(\psi+\frac{\tilde{F}v_\|}{|B|})=\dot{\mathcal{P}_\phi}.
\end{equation}

The statement in the opposite direction is not true: $\dot{\mathcal{P}}_\phi=0$ does not imply the drift velocity in (\ref{vGC}). Indeed, the constancy of $\mathcal{P}_\phi$ can also take into account a toroidal 
 flow: $V=\mathcal{Q} \left[v_\| b+ (m_sv_\|)/(q_s|B|)\nabla \times (v_\|b)\right]+\mathcal{R}(r,\theta)\nabla\phi$, where $\mathcal{R}$ stands for \emph{rotation}. This is because  the equation  (\ref{phimot}) has not been considered yet.  
 What has been shown is a clear correspondence between $\mathcal{P}_\phi$ and the expression of the drift velocity (\ref{vGC}) due to the toroidal symmetry.
 
  When $\psi=R A_\phi$ (\ref{Aphi}) is substituted within $\mathcal{P}_\phi$, it becomes clear also the reason why the gyrokinetic community refers to $\mathcal{P}_\phi$ as the \emph{canonical toroidal momentum}: 
\begin{equation}
\label{pphi}
\mathcal{P}_\phi=\psi+\frac{Fv_\|}{\omega_c} \approx \frac{L_Z}{q_s},
\end{equation}
for $\omega_c$ the cyclotron frequency.

The GC motion is further simplified by requiring the constancy of $w$ and $\mu$. 
It is worth noting that the condition on $w$ is coherent with the neglecting of the electric potential. Indeed in \emph{drift ordering} $A_0$ behaves as $O(\delta)$:
\begin{equation}
\label{w}
w=\frac{v_\|^2}{2}+\mu |B| \approx \mathcal{E};
\end{equation}
however, the present analysis could straightforward include an electric potential if it depends solely on $\psi$.

From $\dot{w}=0$ and $\dot{\mu}=0$, the time derivative of $v_\|$ becomes \cite{brizard09}:
 \begin{equation}
\label{vparaeqmot}
\dot{v}_\|=-\frac{\mu}{v_\|}V \cdot \nabla|B|.
\end{equation}
Now there is a clear correspondence between (\ref{eqmot}) and (\ref{vparaeqmot}) with the following system of equations:
\begin{equation}
\label{eqmotQI} 
\eqalign{\dot{w}&=0,\\ 
\dot{\lambda}&=0,\\
\dot{\mathcal{P}_\phi}&=0.
}
\end{equation}
The GC projected orbit may be defined by the initial conditions: $w=w_0,  \lambda=\lambda_0$ and $\mathcal{P}_\phi=\mathcal{P}_{\phi0}$\footnote{Sometimes the sign of $v_\|$ has to be specified, how it will be clarified in Section \ref{sec:orb}.}. As mentioned above, QIs are the imposed constants $w, \lambda$ and $\mathcal{P}_\phi$.
It becomes clear that a distribution function depending on the QIs  will describe a distribution of GC (projected) orbits. 

When the orbits behavior is considered in a simulation or in a theoretical analysis, it is necessary to take into account the finite orbit width (FOW) effects. It is possible to estimate the relevance of the FOW directly from the expression of $\mathcal{P}_\phi$. Indeed, the term $F v_{\|}/\omega_c$ in (\ref{pphi}) expresses how far apart an orbit will be from the poloidal flux surface coordinate $\psi=\mathcal{P}_{\phi}$: e.g. characteristic \emph{banana} orbits describing trapped particles will have the tip of the banana at $\psi=\mathcal{P}_{\phi}$, because here $v_\|=0$, whilst the banana orbit width will depend on the maximum and minimum reachable values of $v_\|$.

 Several possibilities may arise. Concerning the electrons, $\omega_c$ is big enough. In this case it is possible to describe the electron orbits directly on the flux surface, because $\psi \sim \mathcal{P}_{\phi}$. This is called the small orbit width (SOW) case. If $\omega_c \to \infty$ then $\psi= \mathcal{P}_{\phi}$. This is called the  ideal ZOW (Zero Orbit Width) case and the  equilibrium can be described by a distribution function depending on $(\psi,w,\lambda)$. In the ZOW case, the often used equilibrium distribution function is the local Maxwellian distribution function (where $e$ is for electrons):
\begin{equation}
\label{localM}  
f_M (\psi,w)=\frac{n_e(\psi)}{(2\pi)^{3/2} v_{te}^3(\psi)} e^{- w/v_{te}^2(\psi)},
\end{equation}
where $v_{te}=\sqrt{T_e/m_e}$.
A useful extension of this distribution function, taking into account SOW effects, is depicted by:
\begin{equation}
\label{canoM} 
f_M (\mathcal{P}_\phi,w)=\frac{n_e(\mathcal{P}_\phi)}{(2\pi)^{3/2} v_{te}^3(\mathcal{P}_\phi)} e^{- w/v_{te}^2(\mathcal{P}_\phi)},
\end{equation}
known as \emph{canonical Maxwellian} distribution function \cite{idomura}. The local Maxwellian distribution function can be considered the ZOW limit of the canonical Maxwellian distribution function, appropriate for SOW effects. 
In the following some difficulties, arising specially when large orbit width effects are not negligible, will be described.

\subsection{The current way to describe a gyrokinetic equilibrium with Finite Orbit Width effects}
\label{subsec:problems}
Several difficulties arise when the considered orbit has a large width. Firstly it is analyzed  what happens when the approximation $\psi \sim \mathcal{P}_\phi$ fails.  This is more evident for the passing particles, for which $v_\| \neq 0$ almost always, causing a shift to the whole orbit respect to the flux surface value $\psi=\mathcal{P}_\phi$, as can be easily visualized anticipating the Figure \ref{orbits}(c) where the passing orbit in $(r,\theta)$ coordinates are shown not to intersect the dashed line corresponding to the radius $r_{tip}$, when $\psi =\mathcal{P}_\phi$. The GC poloidal flux surface coordinate $\psi$ may be approximated by its orbit averaged value: $ \langle \psi \rangle_{orb}= \langle\psi \rangle_{orb}(\mathcal{P}_\phi,w,\lambda,\sigma)$. This idea has been given by P. Angelino et al. \cite{angelino}, who also suggest  the estimate $\langle \psi \rangle_{orb} \sim \psi_0= \mathcal{P}_{\phi}- (m_s R_0/q_s) \sigma\sqrt{2w}\sqrt{1-\lambda B_0}\ \mathrm{H}(1/B_0-\lambda)$, where the mass $m_s$ and the charge $q_s$ refer to the examined species, $R_0$ and $B_0$ are the major radius and the magnetic field magnitude at the magnetic axis, $\sigma=\mbox{sgn}(v_\|/v)$ and the \emph{Heaviside function} $ \mathrm{H}(1/B_0-\lambda)$ ensures that the square root is well defined. A more precise estimation will be given in (\ref{psiorb}) or (\ref{psiorb2}) in Section (\ref{subsec:orbave}). In \cite{angelino} it is also suggested to slightly modify the equilibrium distribution function from a canonical Maxwellian to a \emph{biased canonical Maxwellian} (where $b$ means \emph{bulk} and $v_{tb}=\sqrt{T_b/m_b}$):
\begin{equation}
\label{biasedM}  
f_M (\psi_0,w)=\frac{n_b(\psi_0)}{(2\pi)^{3/2} v_{tb}^3(\psi_0)} e^{- w/v_{tb}^2(\psi_0)}.
\end{equation}
It is worth noting an inconvenience: having a density in $\psi_0$ means assigning a given number of GCs with a specific $\psi_0$ value that can be obtained from several $\mathcal{P}_\phi,w,\lambda$ (and $\sigma$). These QIs correspond to different orbits that span a wide portion of the GC configuration space. It becomes quite difficult to initialize that distribution function in a marker loading subroutine in a gyrokinetic code. Obviously this kind of problems are commonly addressed by ignoring the initial distribution function but possibly enhancing the number of markers to reduce the statistical fluctuation noise.

A more serious problem consist in the reproduction of the experimental profiles as for the experimental density profiles $n_{exp}(\psi)$ (for the temperature the analysis is even more complicated). Indeed, the fine property $n_{exp}=n_e$ of the Maxwellian distribution (\ref{localM}) is definitively lost. This happens because $\psi$, a spatial variable, has been substituted by $\psi_0$ that depends also on the GC velocity space variables ($w$ and $\lambda$). The same problem arises with the canonical Maxwellian distribution function because also $\mathcal{P}_\phi$ mixes spatial coordinates with velocity coordinates. The properties of the Maxwellian distribution function are destroyed also considering $\mathcal{P}_\phi$ (or $\psi_0$) as a spatial variable independent from the velocity space variables. Indeed the Jacobian necessary to achieve this independency will anyway break the characteristic Gaussian behavior of the integrand in the velocity coordinates. Neverthless it seems very useful, as will be shown in the next section, to consider   $\mathcal{P}_\phi$ as an independent spatial coordinate\footnote{This is not a conceptual difficulty for those who works in gyrokinetics theory where also the GC spatial position is the difference of the particle spatial position  minus a gyroradius that depends on the perpendicular velocity.}.

Up to now the Maxwellian-like distribution function has been described. It is worthy to analyze other equilibrium (or steady state) distribution functions which are useful also for describing fast particles coming from fusion reactions or from external heating sources. The following model distributions are often used:
the Slowing Down (SD) distribution function for fusion alpha particles 
\begin{equation}
\label{SDdist}
f_{SD} (\psi,w)=\frac{\tau_S S_\alpha(\psi)}{8\sqrt{2}\pi}\frac{\mathrm{H}(w_1-w)}{w^{3/2}+w_c^{3/2}},
\end{equation}
where $\tau_S$ is the \emph{Spitzer SD time} \cite{spitzer} and $w_c=v_c^2/2$ is the \emph{critical energy} \cite{sivukhind,stix} and $S_\alpha$ is the \emph{source} term which corresponds to the density;
the anisotropic SD distribution function for suprathermal ions coming from NNBI (Negative Neutral Beam  Injection) heating \cite{core}
\begin{equation}
\label{NBI}
f_{NNBI} (\psi,w,\xi)=\frac{\tau_S S_D(\psi)}{8\pi\sqrt{2\pi \Delta(\psi)}}\frac{\mathrm{H}(w_1-w)}{w^{3/2}+w_c^{3/2}}e^{-[\xi-\xi_0]^2/2\Delta(\psi)},
\end{equation}
where $\xi=v_\|/v$, $w_1$ is the \emph{beam energy}, $S_D(\psi)$ is the \emph{source} term and $\Delta(\psi)$ gives the spread of the pitch angle distribution centered at $\xi_0$; the single pitch angle ICRH (ion cyclotron radiation heating) distribution function \cite{zonca00} for the minority population 
\begin{equation}
\label{singlpit}
\fl
f_{ICRH} (\psi,w,\lambda)=\frac{n_m(\psi) (r/R_0)^{1/2} \theta_b}{2\pi^2 B_0 v_{tm}^3(\psi) \Gamma(3/4)}\left( \frac{v_{tm}(\psi)}{w}\right)^{3/4} \delta(\lambda-\lambda_0) e^{-w/v_{tm}^2(\psi)},
\end{equation}
where $\theta_b$ is the \emph{bounce} angle near the tip of the \emph{banana} orbit, $v_{tm}=\sqrt{T_m/m_m}$, $\Gamma(z)$ is the \emph{Gamma function} and $n_m(\psi)$ is the density ($m$ stands for minority);  the modified biMaxwelian distribution function \cite{cooper}
\begin{equation}
\label{2M}  
\fl
f_{2M} (\psi,w,\lambda)=n_h(\psi)\left[ \frac{m_h}{2\pi T_\perp(\psi)}\right]^{3/2}  \exp \left\{-m_h w \left[\frac{\lambda B_{res}}{T_\perp(\psi)}+\frac{|1-\lambda B_{res}|}{T_\|(\psi)} \right] \right\} ,
\end{equation}
 useful when the pressure tensor is diagonal but anisotropic ($p_\| \neq p_\perp$) and where $m_h$ is the mass of the considered \emph{hot} species, $B_{res}$ is a resonant magnetic field, $n_h(\psi)$ is the density and the temperatures $T_\perp$ and $T_\|$ can be deduced from the high energy limit respectively when $\lambda B_{res}=1$ and when $\lambda=0$. 
 
The procedure analyzed before may be used also for the following distribution functions: the functional form is preserved and $\psi$ has to be substituted with its orbit averaged value $\langle \psi \rangle_{orb}$ (as done in \cite{fu} for the SD case). A similar prescription has to be used for the other evolving variables as $\xi$ substituted with $\langle \xi \rangle_{orb}$ and so on.
What is guaranteed with this ansatz is the dependency on QIs and the recovering of the commonly used distribution functions in the ZOW case, at least as regarding the dependence within $\psi$. However this is not the only way to proceed. 

In this work it is proposed an alternative construction of the equilibrium distribution function which guarantees the above properties, but also the following ones: $\mathcal{F}_{eq}$ will behave according to a Boltzmann-like parametric distribution function and it will preserve useful integrability properties in the FOW case.
How to build $\mathcal{F}_{eq}$ will be shown in Section \ref{sec:good} whilst in the next section the orbits behavior will be described. Indeed, as mentioned before, $\mathcal{F}_{eq}(\mathcal{P}_\phi,w,\lambda)$ describes a distribution of GC (projected) orbits. The analysis of orbit theory issues will serve to better understand the proposed equilibrium model distribution function. Moreover, in the next section   the second condition (\ref{2}), which is mostly ignored in the simpler theoretical models but useful in more realistic tokamak contexts, will be dealt with.


\section{Orbit theory fundamentals}
\label{sec:orb}
The unperturbed GC projected orbit is easily described with the following set of independent variables: $\psi, \mathcal{P}_\phi, w$ and $\lambda$ (and eventually $\sigma$). This system of reference is singular in the ZOW case, when $\mathcal{P}_\phi=\psi$. However when the orbit behavior is considered the FOW effects must be taken into account. Using the definition of $\mathcal{P}_\phi=\psi+Fv_\|/\omega_c$ and the relation $v_\|^2=2w(1-\lambda |B|)$, it is obtained:
\begin{equation}
(\mathcal{P}_{\phi}-\psi)^2=\frac{2w\tilde{F}^2(1-\lambda|B|)}{B^2} \mbox{ , with       } \tilde{F}=m_sF/q_s. 
\end{equation}
Multiplyng both sides with $B^2$, the unique positive solution of the second order equation in $|B|$ is denoted by $B_{orb}$:
\begin{equation}
\label{Borb}
B_{orb}=\frac{w\lambda\tilde{F}^2}{(\mathcal{P}_{\phi}-\psi)^2}\left\{ \left[1+\frac{2(\mathcal{P}_{\phi}-\psi)^2}{w\lambda^2\tilde{F}^2}\right]^{1/2}-1\right\}.
\end{equation}
$B_{orb}$ is the intensity of the magnetic field magnitude $|B|$  seen from the GC along its orbital motion. It is worth noting that $B_{orb}$ depends only by $\psi,\mathcal{P}_\phi, w$ and $\lambda$ which substantiates this choice of orbit coordinates. Once the ($r,\theta$) map of the magnitude of the magnetic field $|B|(r,\theta)$ is known, it is possible to describe the projected orbit in poloidal coordinates from the implicit relation:$|B|(r,\theta)=B_{orb}(\psi(r),\mathcal{P}_\phi, w$,$\lambda)$.

 As an example, it is considered one of the most analyzed model \cite{shafranov1} of tokamak plasma with nested circular flux surfaces with a Shafranov shift $\Delta(r)$ and a little inverse aspect ratio $\varepsilon$, described with \emph{Shafranov coordinates} (see Appendix A.2), when $\Delta^\prime=\mathcal{O}(\varepsilon)$ (\ref{B2}, \ref{Bmagnitude}):  
\begin{equation}
\label{Bshaf}
\eqalign{
 B&=\frac{rF[1+\mathcal{O}(\epsilon^2)]}{qR(R_0-\Delta)(1-\Delta' \cos \theta)} e_\theta+ \frac{F}{R} e_\phi \rightarrow \\
&\rightarrow |B|=\frac{F}{R}\left[ 1+\frac{r^2}{2q^2(R_0-\Delta)^2}+\mathcal{O}(\varepsilon^3)\right], 
}
\end{equation}
where $e_\theta=(\nabla \psi \times \nabla \phi) /|\nabla \psi \times \nabla \phi|$ is the poloidal unit vector and $q(r)$ is the safety factor.
From (\ref{Bshaf}) with $B_{orb}$ in place of $|B|$ and from $R=R_0-\Delta(r)+r\cos \theta$, it is possible to express $\cos \theta$ as a function of $\psi,\mathcal{P}_\phi, w$ and $\lambda$:
\begin{equation}
\label{orbitpol}
\cos \theta = \frac{F}{rB_{orb}}\left[ 1+\frac{r^2}{2q^2(R_0-\Delta)^2}+\mathcal{O}(\varepsilon^3)\right]+\frac{\Delta-R_0}{r}.
\end{equation}
Known the functions $\psi(r), \Delta(r), q(r), F(\psi)$ and given $\mathcal{P}_\phi, w$ and $\lambda$,(\ref{orbitpol}) is the orbit expressed in poloidal coordinates ($r,\theta$).

 Returning to the general $|B|(r,\theta)$ case, it is possible to plot the orbit projection in the ($r,\theta$) poloidal reference system if the QIs are assigned.
It becomes easy to classify the GC orbits, also thanks to the chosen $\lambda$ coordinate. Indeed, the following relation is obtained equating $B_{orb}$ with $|B|$ and substituting $w\lambda^2$ with $\chi$ in (\ref{Borb}):
\begin{equation}
\label{Lambda}
\fl
\lambda=\frac{\chi\tilde{F}^2}{[\mathcal{P}_{\phi}-\psi(r)]^2|B|(r,\theta)}\left\{ \left\{1+\frac{2[\mathcal{P}_{\phi}-\psi(r)]^2}{\chi\tilde{F}^2}\right\}^{1/2}-1\right\}=\Lambda(r,\theta;\mathcal{P}_{\phi},\chi).
\end{equation} 
No matter how complex it can be the map $|B|(r,\theta)$, provided that the magnetic flux surfaces does exist, the $\Lambda(r,\theta)$ surface defined in (\ref{Lambda}) plays the same role of the potential energy in the classification of orbits in mechanics: the analysis which is based on the stationary points of the potential energy.

\begin{figure}[htbp]
\begin{center}
\mbox{
\includegraphics[width=6.5cm]{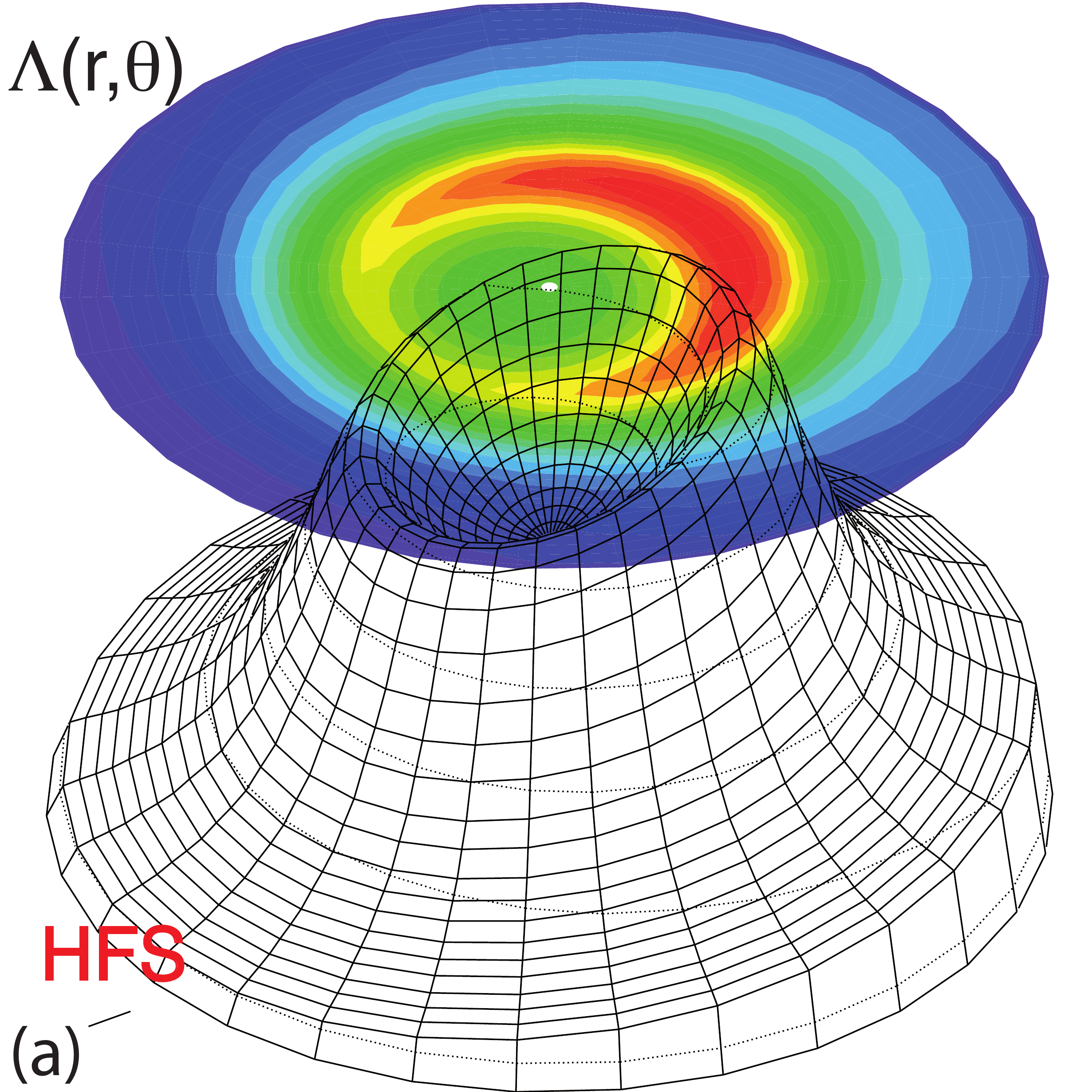}$\;\;\;$
\includegraphics[width=7.5cm]{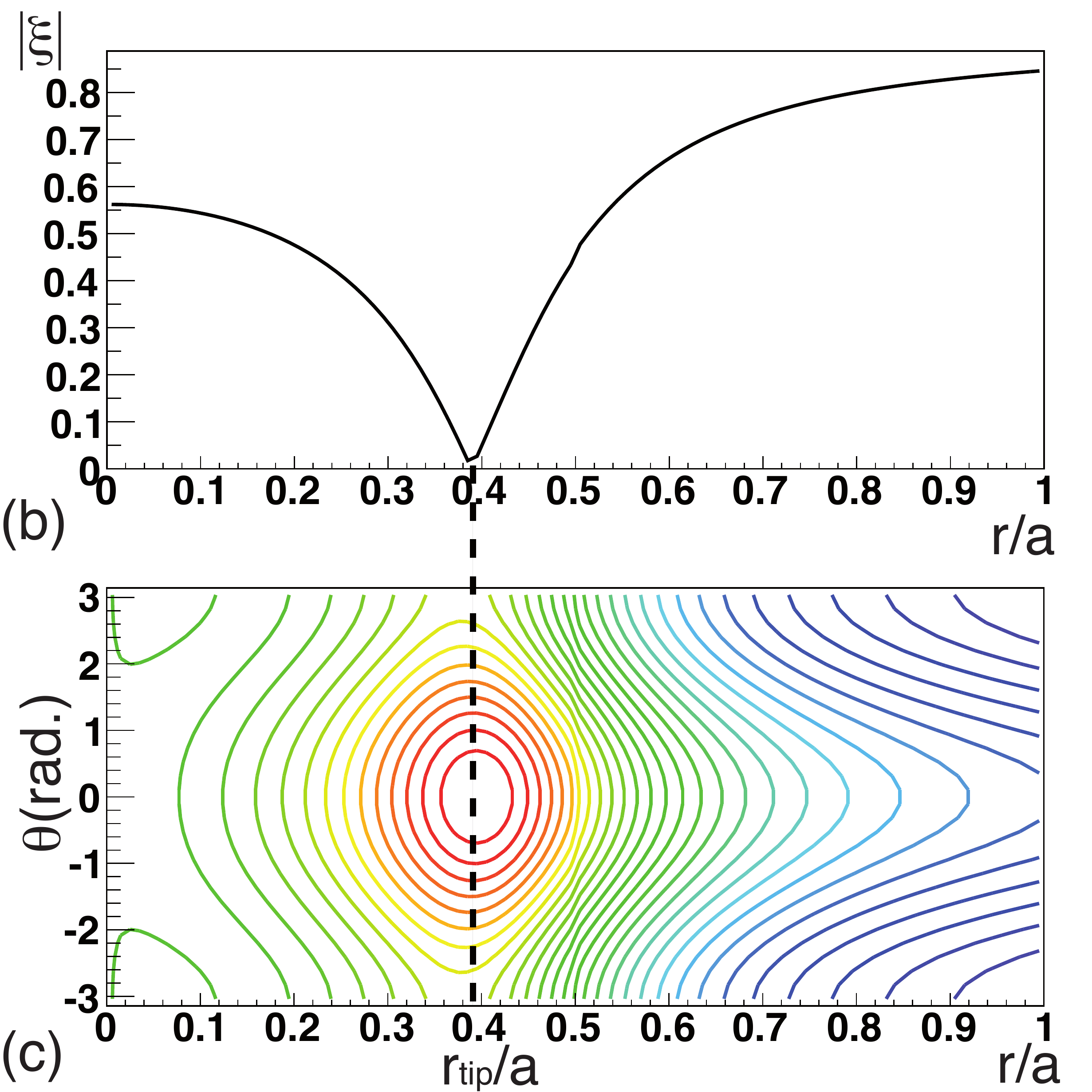}
}
\caption{{\bf (a)} Surface Polar Plot of $\Lambda(r,\theta)$ defined in (\ref{Lambda}), at $\mathcal{P}_\phi$ and $\chi \equiv w\lambda^2$ fixed. The (projected) orbits are the level curves of  $\Lambda$ also recognizable in the Contour Plot just above the surface. As for example, the banana shape of the banana orbits is easily recognized.
{\bf (b)}$|\xi| \equiv |v_\||/v$ versus $r/a$ for the same case considered in the left figure. When $|\xi|=0$ the GC approximately reverse its motion, in which case the orbit is considered trapped. {\bf (c)} The projected orbits plotted on the $r/a-\theta$ plane are level curves of the $\Lambda$ surface (on the left). It is possible to deduce the \emph{tip} coordinates $(r_{tip},\theta_{tip})$ for each of the visualized trapped orbits from the intersections of those orbits with the dashed vertical line corresponding to $\xi=0$.}
\label{orbits}
\end{center}
\end{figure}

Given a  simple map $|B|(r,\theta)$ as (\ref{Bshaf}) can be, and fixing $\mathcal{P}_\phi$ and $\chi$, the surface $\Lambda$ is plotted in the polar ($r,\theta$) reference system in  Figure {\ref{orbits}}(a). Assigning a value to $\lambda$ means cutting horizontally the $\Lambda$ surface. The GC projected orbits are recognized as the level curves of $\Lambda$ as clarified in Figure {\ref{orbits}}(c). In the depicted case it is assumed an up-down symmetry for simplicity. From the shown level curves of $\Lambda$ it is possible to  distinguish the \emph{lost} orbits, if they intersect the radius $r=r(\psi_a)\equiv a$, where $\psi_a$ is the poloidal flux of the magnetic field for the last nested magnetic flux surface (as for the \emph{separatrix} magnetic flux surface when a \emph{divertor} is present). In the depicted curves, it is possible to qualitatively distinguish the \emph{trapped} orbit  if they make a loop, from the others: the \emph{passing} orbits, usually going from $-\pi$ to $\pi$, or \emph{viceversa}. 

It can be also possible to quantify the above graphical method. For classifying orbits, the most common 
practice is to analyze the sign of the variable $\xi=v_\|/v=\sigma\sqrt{1-\lambda B_{orb}}$ along the orbit: if $\xi$ changes its sign then it admits a zero and the GC orbit will be almost trapped, otherwise it will be almost passing. These differences are depicted on the right of Figure 1 comparing (b) with (c); Figure 1(b) shows the absolute value of $\xi$:
\begin{equation}
|\xi|=\left\{1-\frac{\chi\tilde{F}^2}{[\mathcal{P}_{\phi}-\psi(r)]^2}\left\{ \left\{1+\frac{2[\mathcal{P}_{\phi}-\psi(r)]^2}{\chi\tilde{F}^2}\right\}^{1/2}-1\right\}\right\}^{1/2}.
\end{equation}
  The radius $r_{tip}$ and the angle $\theta_{tip}$ are the coordinates corresponding to $\xi=0$ at given $\mathcal{P}_\phi$ and $\chi$. When $\xi=0$ then $r_{tip}=r(\psi)\mid_{\psi=\mathcal{P}_\phi}$. Depending on $\lambda$, it is possible to obtain or not, as the case may be, the intersection of the orbit with the $r_{tip}$ value (dashed line in Figure 1(c)). When the intersection happens, $\theta_{tip}$ can be evaluated from the implicit relation: $\lambda=\Lambda(r_{tip},\theta_{tip};\mathcal{P}_\phi,\chi)$. While $\theta_{tip}$ depends on $\mathcal{P}_\phi,\chi$ and $\lambda$, it is important to emphasize that $r_{tip}$ depends only on $\mathcal{P}_\phi$,  showing a degeneracy on $\lambda$ and $w$ (or $\chi$). In the  Figure {\ref{orbits}}(c), it is also clear that the \emph{bounce} coordinates $(r_b, \theta_b)$, defined as the values corresponding to $\theta^\prime=0$, slightly differ to the  \emph{tip} coordinates $(r_{tip}, \theta_{tip})$\footnote{The difference between \emph{bounce} and \emph{tip} coordinates can be traced back to depend on the $\nabla |B|$ drift.}. Although $(r_b, \theta_b)\approx (r_{tip}, \theta_{tip})$ is a good approximation, it is possible to be more precise. The classification of the orbit topology will be derived here directly from the geometry of the $\Lambda$ surface.
  \begin{figure}[htbp]
\begin{center}

\includegraphics[width=15cm]{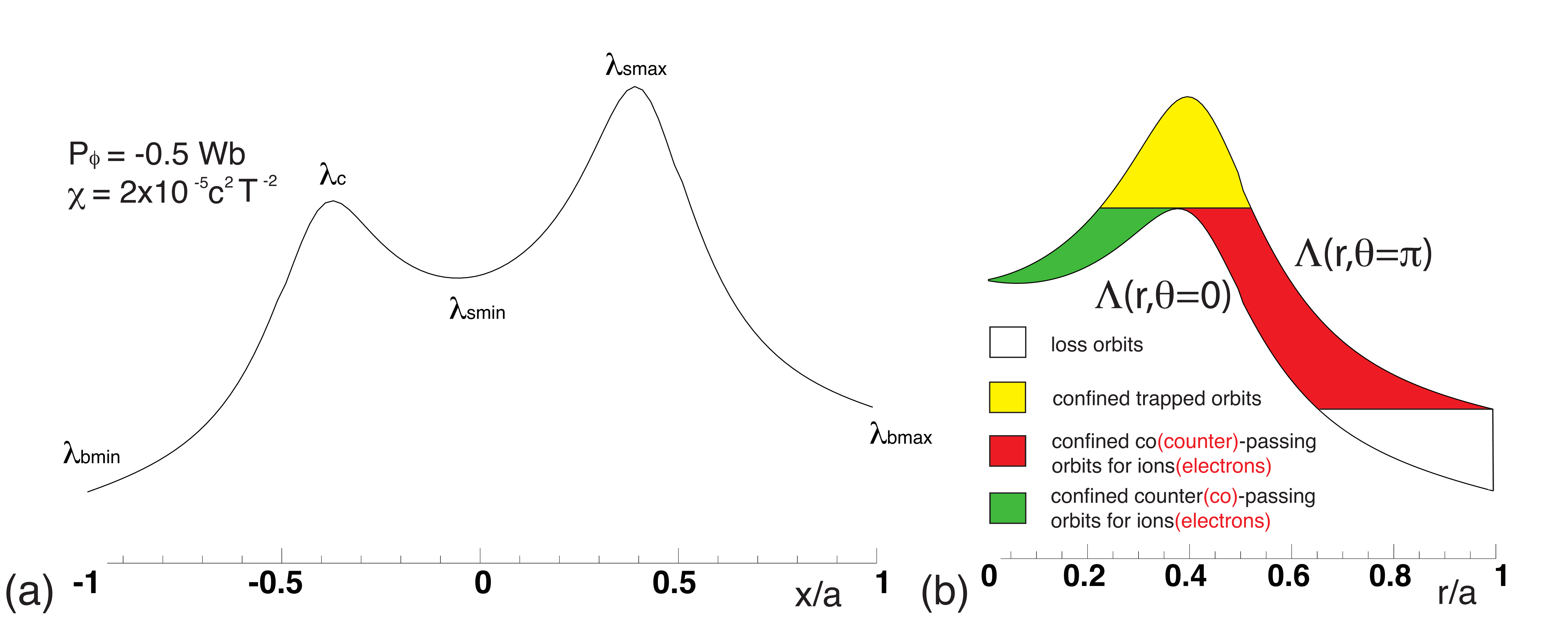}
\caption{{\bf (a)} Equatorial section of $\Lambda$ in Figure \ref{orbits}, in order to illustrate the $\lambda$ values used for the orbits classification: $\lambda_{bmin},\lambda_{bmax},\lambda_{smin},\lambda_{smax},\lambda_c$, as described in the text.{\bf (b)} $(\lambda,r)$ domain of variability subdivided in four zones: the \emph{loss} orbits are enclosed in the white zone, the co(counter)-passing orbits for ions(electrons) are in the red(dark grey) zone, the trapped orbits are in the yellow(off-white) zone and the counter(co)-passing orbits for ions(electrons) are in the green(light grey) zone.}
\label{loss0}
\end{center}
\end{figure}

Generally, at most only five values of $\Lambda$ can determine the whole classification of the orbits, as can be recognized in Figure \ref{orbits}(a): $\lambda_{bmin}$ and $\lambda_{bmax}$ are respectively the minimum and the maximum value of $\Lambda$ at the boundary, when $r=a$. $\lambda_{smin},\lambda_{smax}$ are the local minimum and the local maximum of $\Lambda$, and they are commonly defined as the stagnation points, $\lambda_c$ is the critical value corresponding to the saddle of $\Lambda$. Sometimes, the \emph{critical} orbit (when $\lambda=\lambda_c$) is called the pinch-orbit. 
The coordinates $(r_c,\theta_c)$ are the solution of $\lambda_c=\Lambda(r_c,\theta_c;\mathcal{P}_\phi,\chi)$ when $\Lambda(r,\theta)$ shows the saddle point.

 These particular values of $\Lambda$ depend on $\mathcal{P}_\phi$ and on $\chi$. When an up-down symmetry is considered, they are displaced along the equator characterized by the abscissa $x\in(-a,a)$ . In Figure \ref{loss0}(a), the $\Lambda$ section along the equator have been plotted to better visualize $\lambda_{bmin}, \lambda_{bmax}, \lambda_{smin}, \lambda_{smax}$ and $\lambda_c$ for the same values of $\mathcal{P}_\phi$ and $\chi$ chosen for the plots in Figure \ref{orbits}. It can be seen how $\lambda_{smin}$ and  $\lambda_c$ lie on the High Field Side (HFS) when $\theta=\pi$ while $\lambda_{smax}$ lies on the Low Field Side (LFS) when $\theta=0$.
 
  The Figure \ref{loss0}(b) shows the same $\Lambda$ section, once it has been folded around the $x=0$ axis. The area enclosed between the two branches $\Lambda(r,\theta=0),\Lambda(r,\theta=\pi)$ and the vertical line $r=a$ is the $(\lambda,r)$ domain of variability.
The projected orbits are horizontal lines which connect the boundary of that domain.
Now, it is possible to distinguish unambiguously the following four classes of orbits:(i) the \emph{loss}  orbits are those that touch the vertical line $r = a$, and are indicated in white. For the considered case, this happens for $\lambda \leq \lambda_{bmax}$, on the contrary, the confined orbits have $\lambda > \lambda_{bmax}$(zones with colors). The confined orbits can be divided in  (ii) \emph{trapped} orbits if $\lambda>\lambda_c$, indicated in yellow (off-white for b/w copy), or \emph{passing} orbits if $\lambda<\lambda_c$. The passing orbits are further subdivided between (iii) those with a radius $r< r_c$, the \emph{counter-passing(co-passing)} for ions(electrons), indicated in green (light grey for b/w copy), and (iv) those with a radius $r> r_c$, the \emph{co-passing(counter-passing)} for ions(electrons), indicated in red (dark grey for b/w copy).
 
 \begin{figure}[htbp]
\begin{center}
\mbox{
\includegraphics[width=5cm]{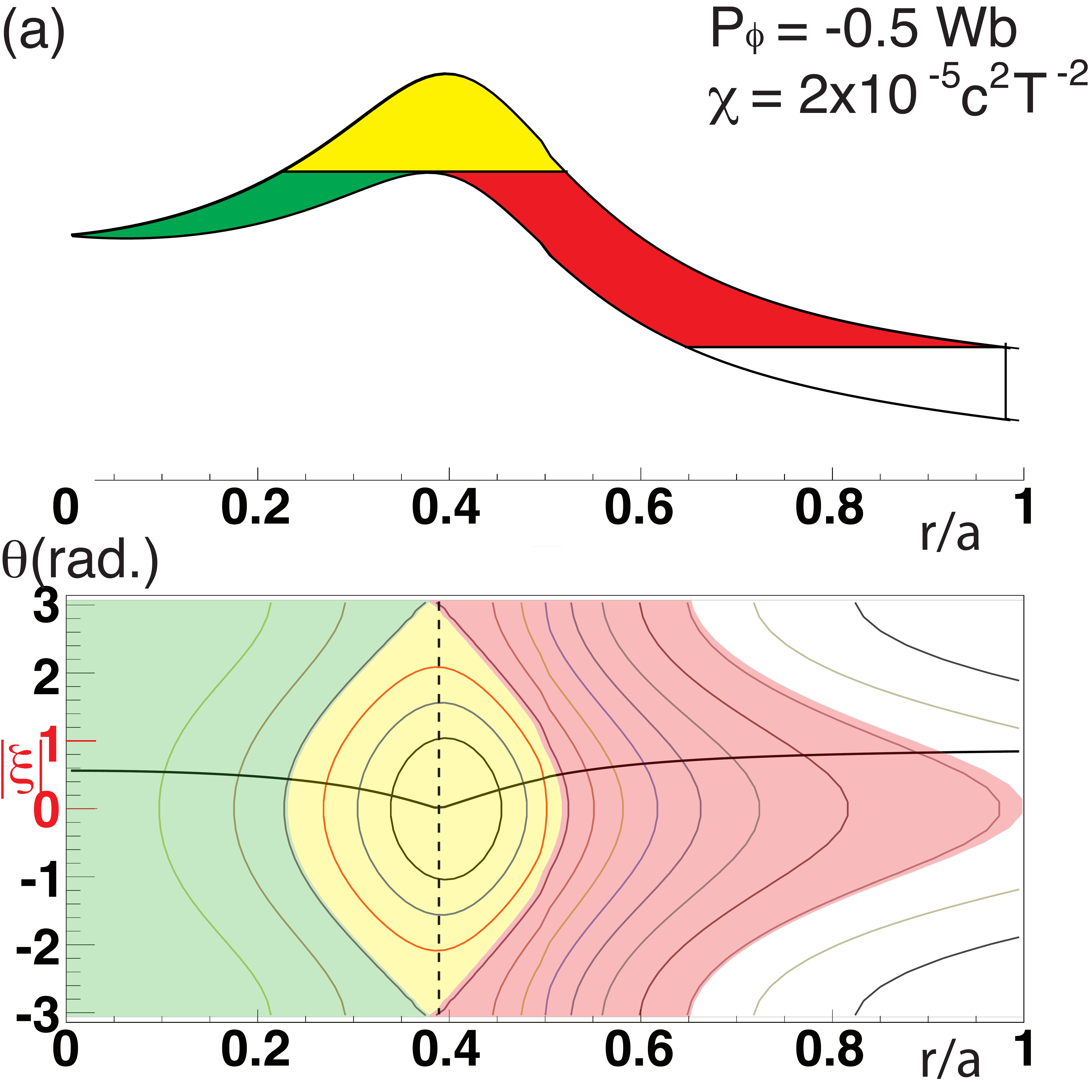}
\includegraphics[width=5cm]{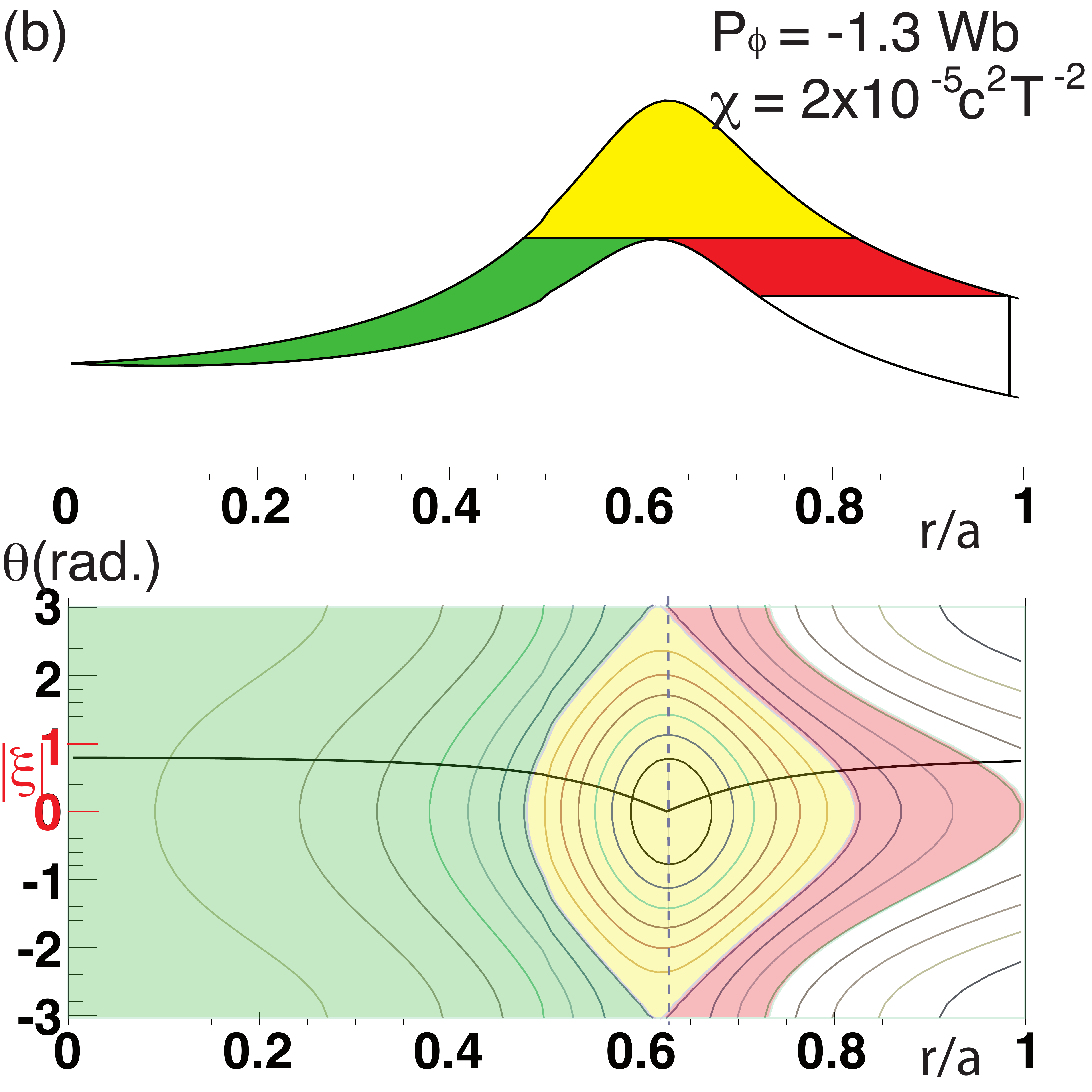}
\includegraphics[width=5cm]{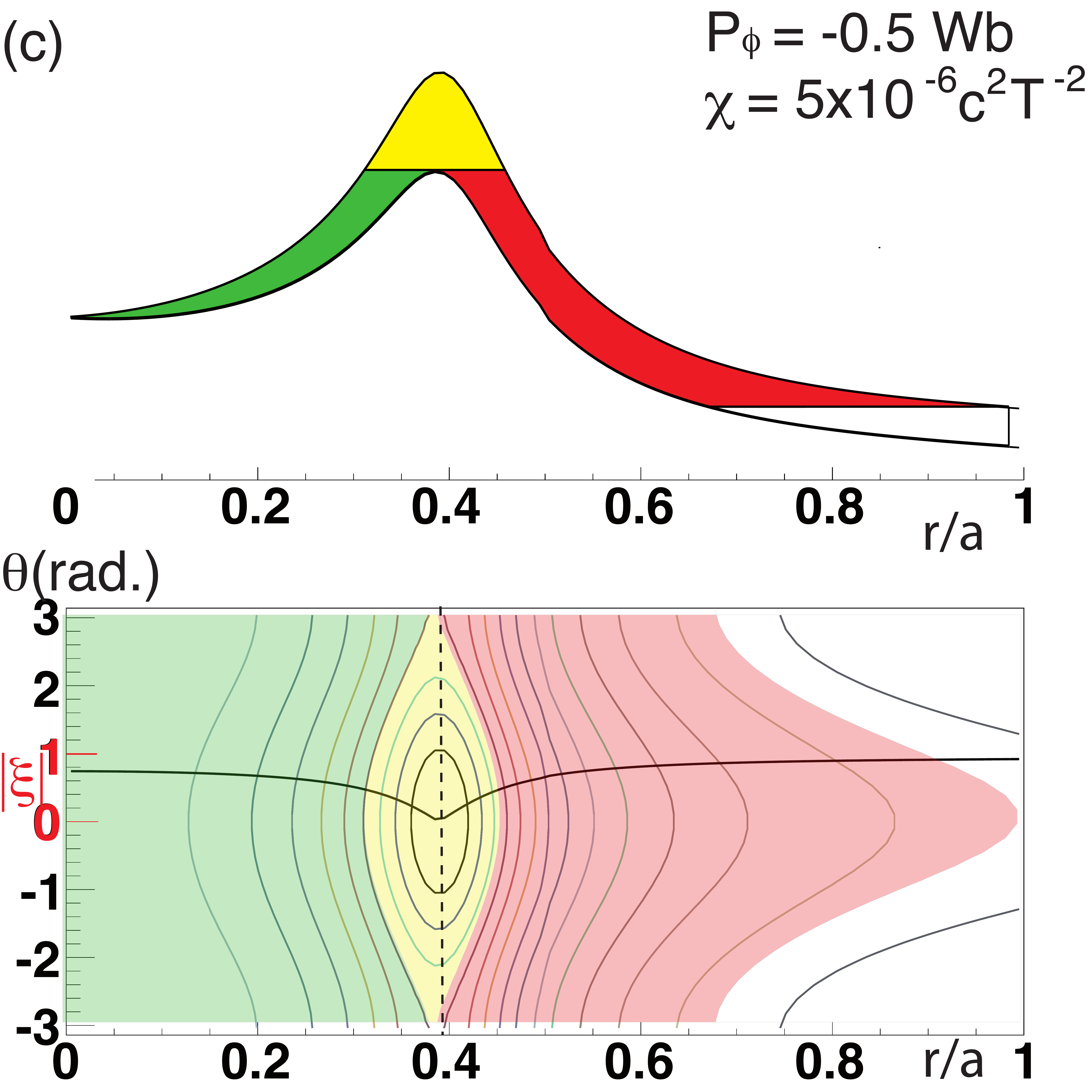}
}
\caption{$(\lambda,r)$ domain of variability (as in Figure \ref{loss0}(b)) above the corresponding orbits plotted on the $r/a-\theta$ plane as level curves of the $\Lambda$ surface (as in Figure \ref{orbits}(c)), with the same color code used in Figure \ref{loss0}(b). In overlapping it is plotted $|\xi|$ as in Figure \ref{orbits}(b) and the $r_{tip}$ value (dashed line). The employed $\mathcal{P}_\phi,\chi$ values are respectively: {\bf (a)} $\mathcal{P}_\phi=-0.5$ Wb and $\chi=2\times10^{-5}\mbox{ c}^2\mbox{ T}^{-2}$, {\bf (b)} $\mathcal{P}_\phi=-1.3$ Wb and $\chi=2\times10^{-5}\mbox{ c}^2\mbox{ T}^{-2}$ and {\bf (c)} $\mathcal{P}_\phi=-0.5$ Wb and $\chi=2\times10^{-6}\mbox{ c}^2\mbox{ T}^{-2}$.}

\label{loss1}
\end{center}
\end{figure}

  Figure \ref{loss1}(a) reproduce the same plot of Figure \ref{loss0}(b) together with the projected orbits plot of Figure \ref{orbits}(c), to better comprehend the relation between the classifications of orbits in the $(\lambda,r)$ domain and the shape of the corresponding orbits in the real space. The real space has been colored indicating the four zones: counter-passing(co-passing) for ions(electrons) in green (light grey), trapped orbits in yellow (off-white), co-passing(counter-passing) for ions(electrons) in red (dark grey) and loss  orbits in white. Those zones are separated by the two branches of the critical orbit $\lambda=\lambda_c$ and by the confined boundary orbit $\lambda=\lambda_{bmax}$.  In addition, to better appreciate the accuracy of the commonly used method based on the sign of $\xi$, it is also plotted the value of $|\xi|$ (same plot of Figure 1(b)) and the dashed vertical line corresponds to $r_{tip}$.
  
  The plots that follow are realized in the same manner as for Figure \ref{loss1}(a), but $\Lambda$ is now obtained with different values of   $\mathcal{P}_\phi$ and $\chi$. In Figure \ref{loss1}(b), $\mathcal{P}_\phi$ is lowered. As a consequence the trapped orbits zone is shifted on the right ($r_c$ is  increased as a consequence of taking $\psi^\prime$ negative). In Figure \ref{loss1}(c), $\chi$ have been lowered and $\mathcal{P}_\phi$ retained as in Figure \ref{loss1}(a), to show how the width of the orbits depends on $\chi$: the orbit width increases or decreases together with $\chi$.  
  
\begin{figure}[htbp]
\begin{center}
\mbox{
\includegraphics[width=5cm]{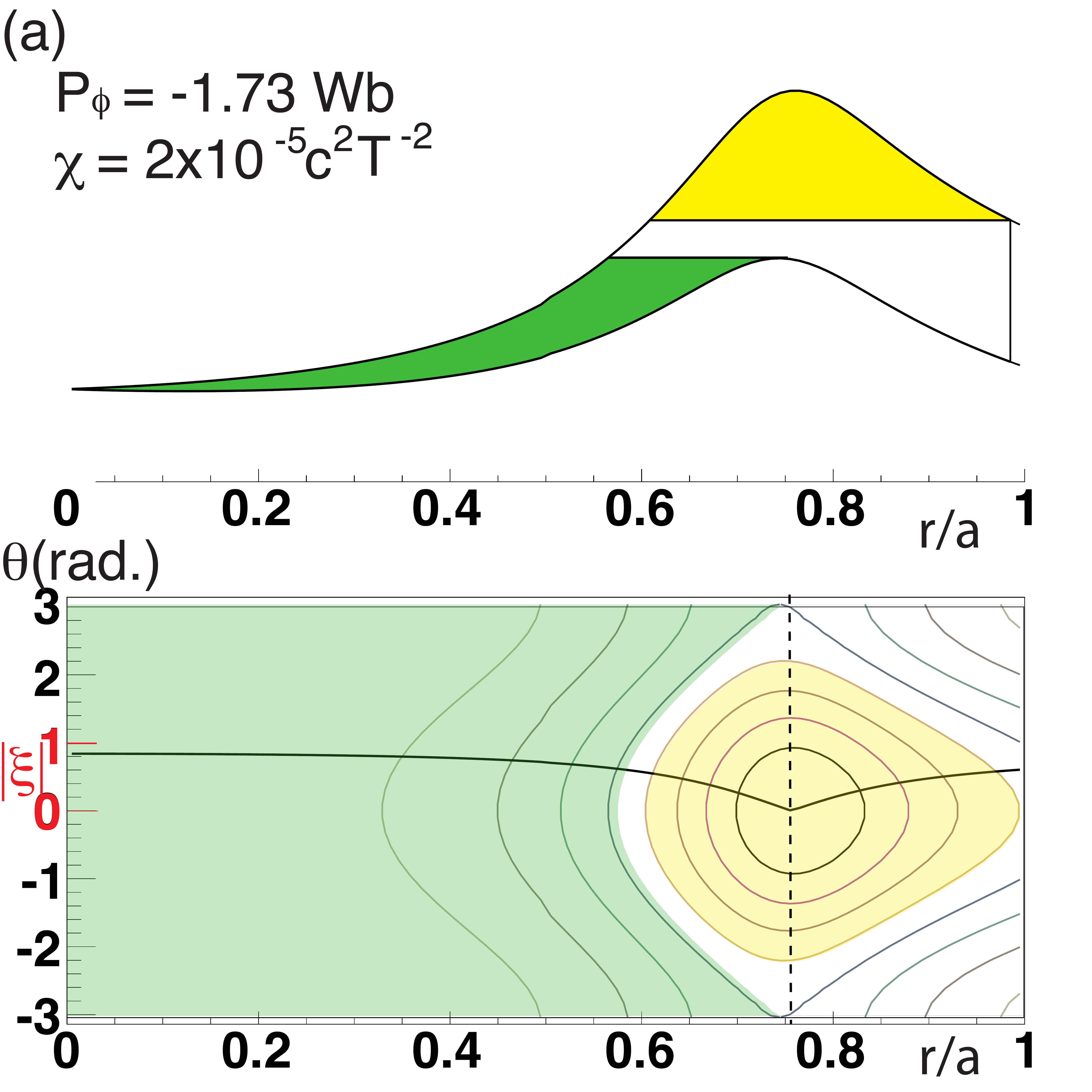}
\includegraphics[width=5cm]{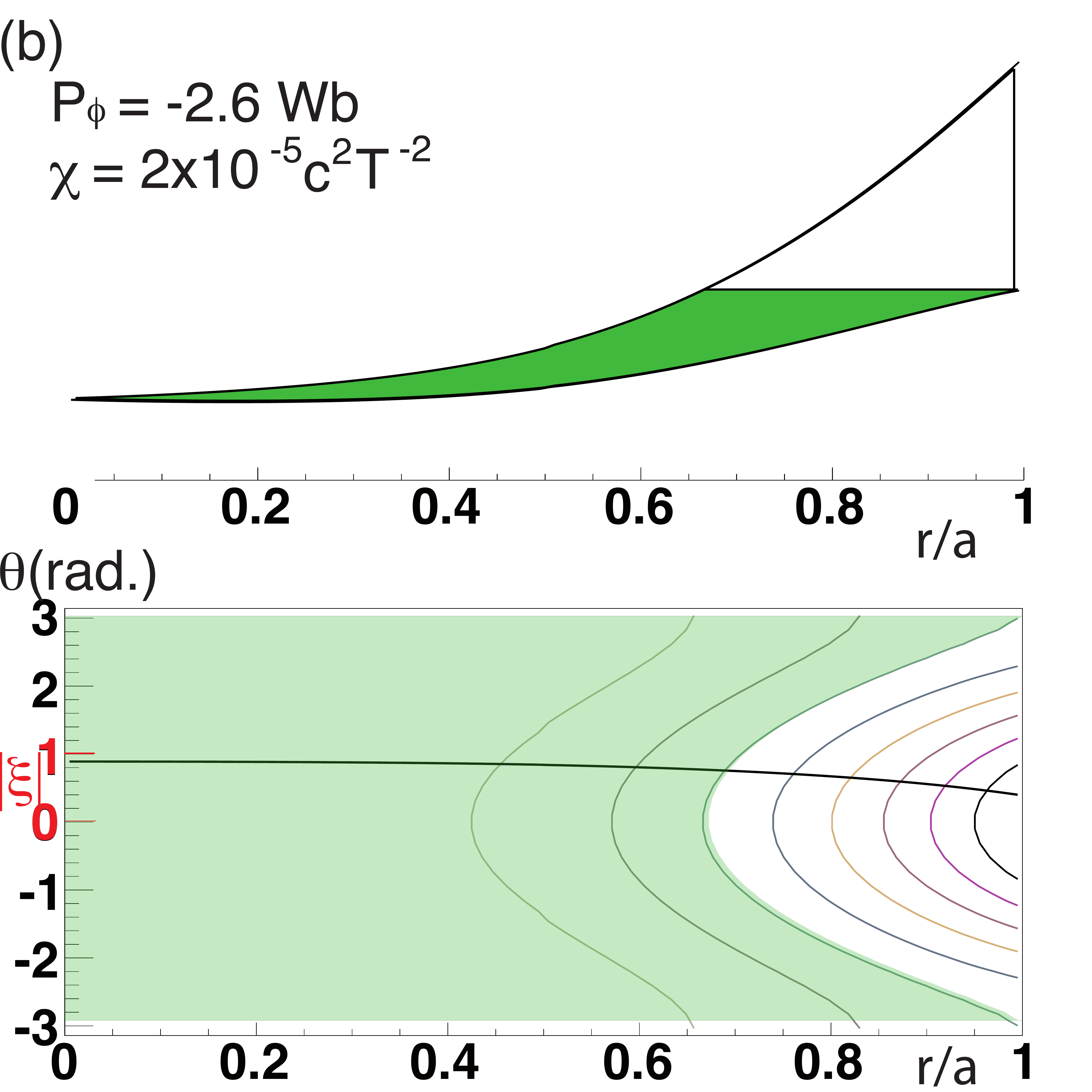}
\includegraphics[width=5cm]{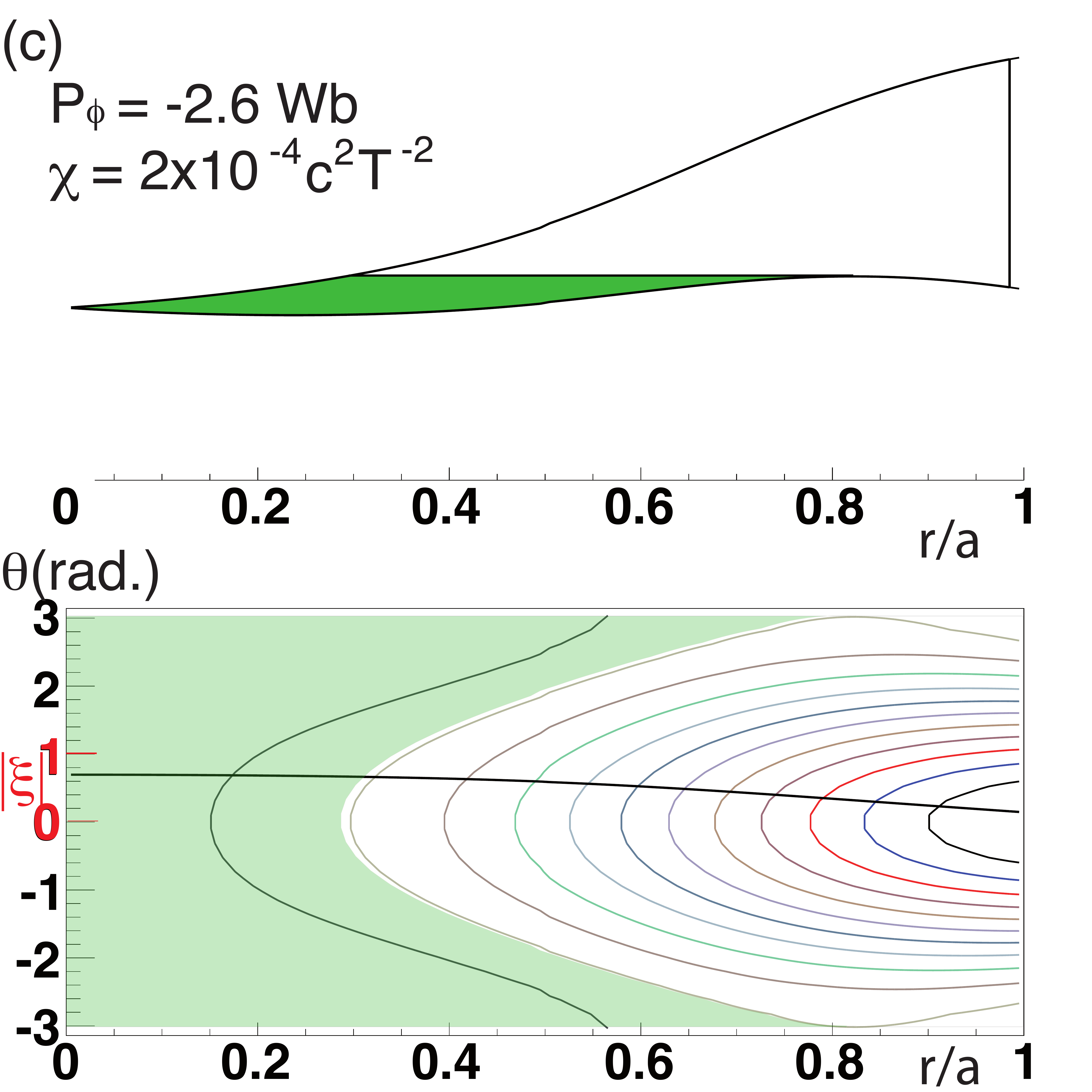}
}
\caption{Same as Figure \ref{loss1} to show the case where confined co-passing orbits for ions (counter-passing orbits for electrons) are not allowed. The employed $\mathcal{P}_\phi,\chi$ values are respectively: {\bf (a)} $\mathcal{P}_\phi=-1.73$ Wb and $\chi=2\times10^{-5}\mbox{ c}^2\mbox{ T}^{-2}$, {\bf (b)} $\mathcal{P}_\phi=-2.6$ Wb and $\chi=2\times10^{-5}\mbox{ c}^2\mbox{ T}^{-2}$ and {\bf (c)} $\mathcal{P}_\phi=-2.6$ Wb and $\chi=2\times10^{-4}\mbox{ c}^2\mbox{ T}^{-2}$.}
\label{loss2}
\end{center}
\end{figure}

  In Figure \ref{loss2}(a), $\chi$ is kept as in Figure \ref{loss1}(b), but $\mathcal{P}_\phi$ is further lowered with the consequence that part of the trapped orbits are now lost: the confined zone becomes disconnected, separated by all the orbits with $\lambda_c \leq \lambda \leq \lambda_{bmax}$.
  
  When  $\mathcal{P}_\phi$ is decreased below $\psi_a$, as depicted in Figure \ref{loss2}(b) and \ref{loss2}(c), there is place only for co-passing ions (or counter-passing electrons). The true reason is that $\Lambda$ doesn't now show any stationary local maximum: $\lambda_{smax}$ doesn't exist in the plasma volume. In Figure \ref{loss2}(c), $\chi$ is increased to show how $r_c$ can differ from $r_{tip}$, indeed, in the present case, $r_{tip}$ doesn't exist because $\xi$ doesn't vanish. Whilst $r_{tip}$ depends only on $\mathcal{P}_\phi$, $r_c$ depends also on $\chi$.
  
\begin{figure}[htbp]
\begin{center}
\mbox{
\includegraphics[width=5cm]{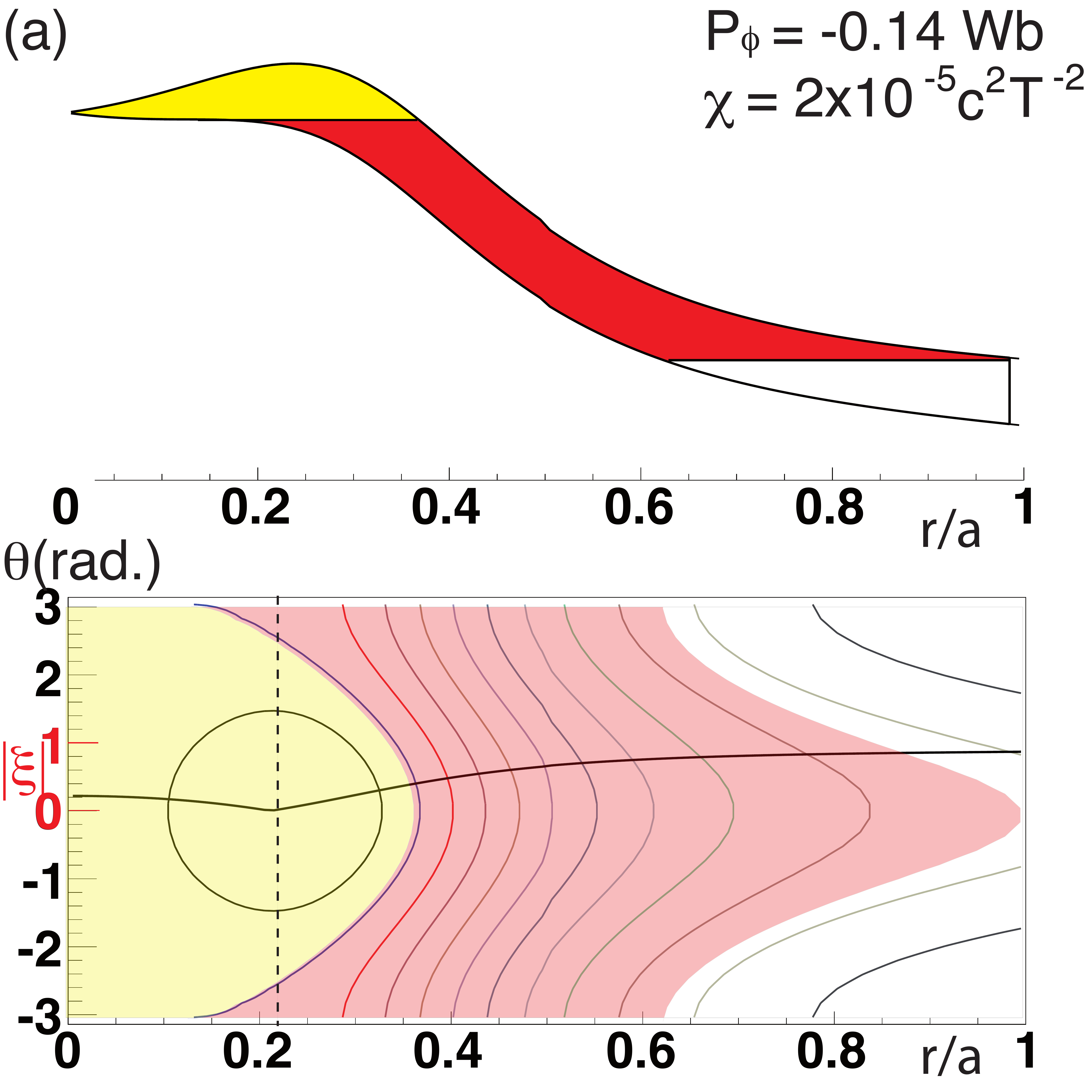}
\includegraphics[width=5cm]{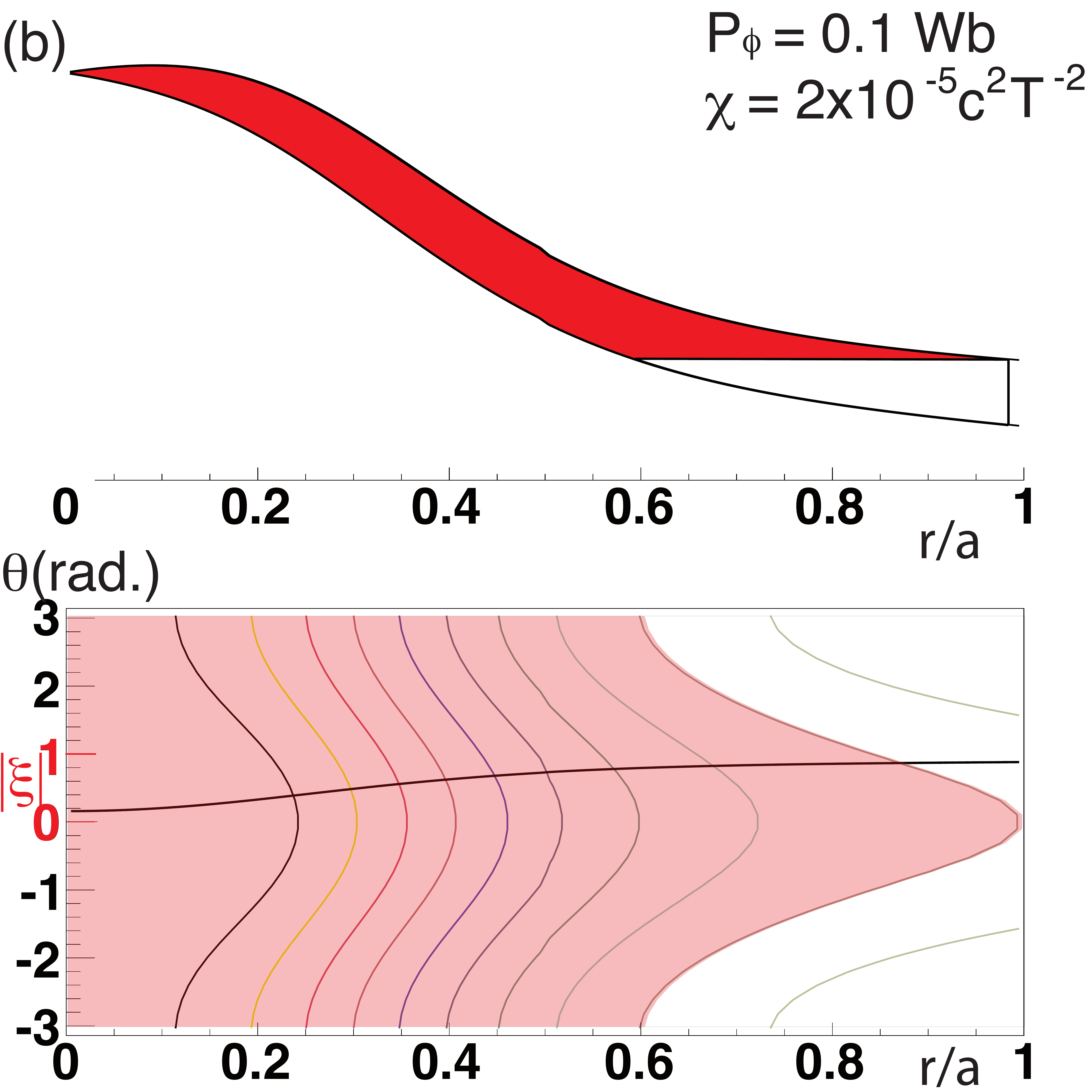}
\includegraphics[width=5cm]{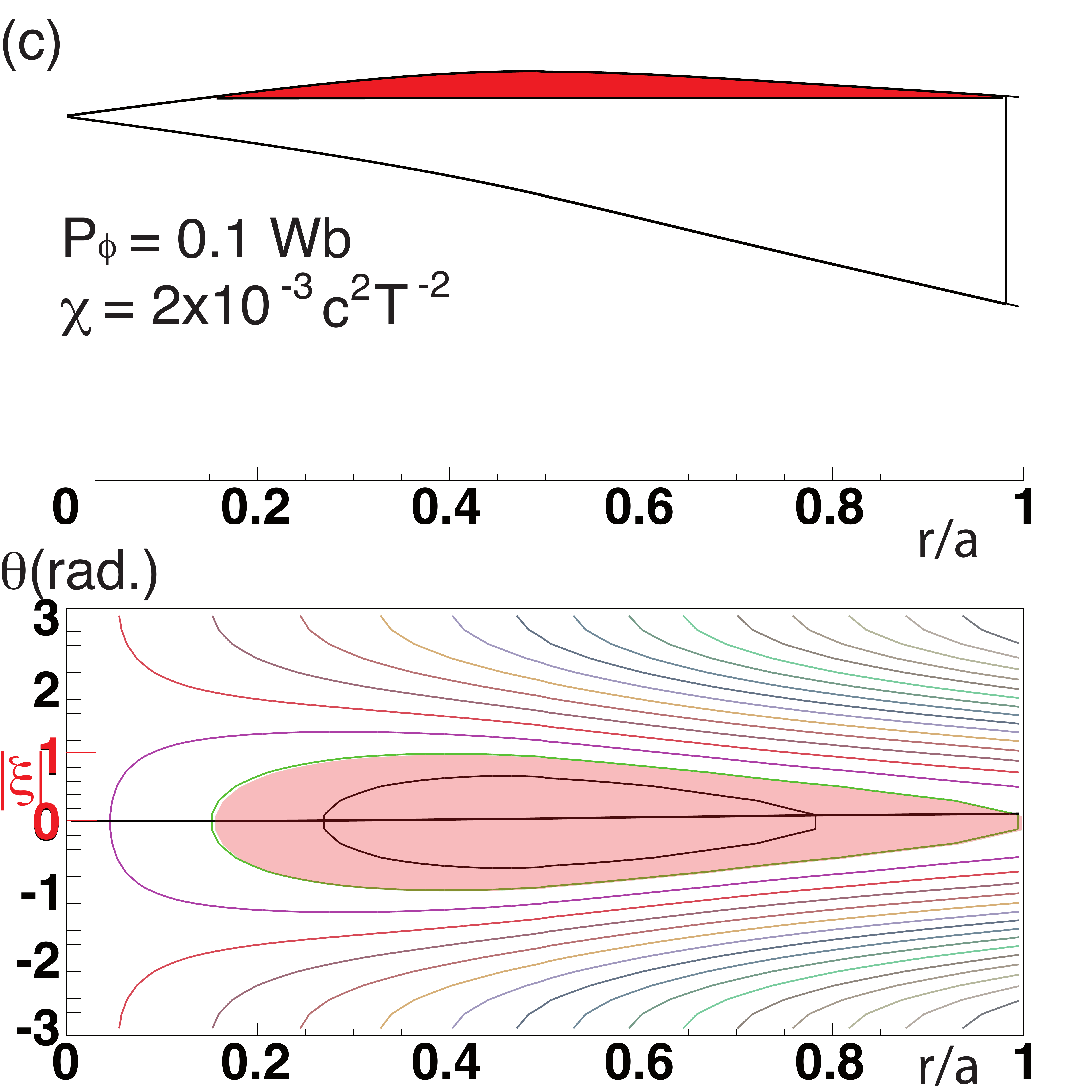}
}
\caption{Same as Figure \ref{loss1} and Figure \ref{loss2} to show the case when confined counter-passing orbits for ions (co-passing orbits for electrons)  are not allowed. The employed $\mathcal{P}_\phi,\chi$values are respectively: {\bf (a)} $\mathcal{P}_\phi=-0.14$ Wb and $\chi=2\times10^{-5}\mbox{ c}^2\mbox{ T}^{-2}$, {\bf (b)} $\mathcal{P}_\phi=0.1$ Wb and $\chi=2\times10^{-5}\mbox{ c}^2\mbox{ T}^{-2}$ and {\bf (c)} $\mathcal{P}_\phi=0.1$ Wb and $\chi=2\times10^{-3}\mbox{ c}^2\mbox{ T}^{-2}$.}
\label{loss3}
\end{center}
\end{figure}

  In Figure \ref{loss3}(a), $\chi$ is kept as in Figure \ref{loss1}(a), but $\mathcal{P}_\phi$ is increased until $\lambda_{smin}$ is absent. The critical orbit has only one branch which is shown to separate the trapped orbits from  the co-passing ions (counter-passing electrons). There is no place for counter-passing ions (co-passing electrons) because there aren't orbits with $\lambda<\lambda_c$ and $r<r_c$.  In the present case, the difference between $r_c$ and $r_{tip}$  is such that the common method of orbits classification, based on the sign of $v_\|$, cannot be applied without discrepancies.
  
  When $\mathcal{P}_\phi>\psi(r)\mid_{r=0}=0$ the $\Lambda$ geometry changes further: $\lambda_{smax}$ is the only stationary point. In this case all the confined orbits are considered co(counter)-passing for ions(electrons). The Figures  \ref{loss3}(b) and (c) differ only on the $\chi$ value but they represent a similar case.  
  

\subsection{Orbit average, characteristic frequencies and second invariant}
\label{subsec:orbave}
In the orbit coordinates $\psi, \mathcal{P}_\phi, w$ and $\lambda$ (and eventually $\sigma$) it is possible to express the orbit average. The orbit average of a quantity $A$ is defined as a time average along the path $C$ which is the closed projected orbit (only the projection of the orbit is always closed): 
\begin{equation}
\label{orbavecl}
\langle A \rangle_{orb}=\frac{\int_{C} A\, dt}{\int_{C} \, dt}=\frac{\int_{C} A\, d\theta/\dot{\theta}}{\int_{C} \, d\theta/\dot{\theta}}=\frac{\int_{C} A\, dr/\dot{r}}{\int_{C} \, dr/\dot{r}}=\frac{\int_{C} A\, d\phi/\dot{\phi}}{\int_{C} \, d\phi/\dot{\phi}}.
\end{equation}
It is common practice to substitute one of the relations (\ref{eqmot}) or (\ref{phimot}) to evaluate the integral. An alternative procedure is here adopted which uses the relations (\ref{eqmotQI}) expressing the QIs. Thus, the QI average of $A$ is defined as follows:
\begin{eqnarray}
\label{QIave}
\langle A \rangle_{QI}=\frac{\int A \, \mathcal{Q}^{-1} \delta(\tilde{\mathcal{P}}_{\phi}-\mathcal{P}_{\phi})\delta(\tilde{w}-w)\delta(\tilde{\lambda}-\lambda)\, d^3\tilde{x}d^3\tilde{v}}{\int  \, \mathcal{Q}^{-1}\delta(\tilde{\mathcal{P}}_{\phi}-\mathcal{P}_{\phi})\delta(\tilde{w}-w)\delta(\tilde{\lambda}-\lambda)\, d^3\tilde{x}d^3\tilde{v}},
\end{eqnarray}
where the integral is evaluated on the whole phase space when the values of $\mathcal{P}_\phi$, $w$ and $\lambda$ uniquely determine the orbit. When there is a degeneracy, as in the case of the co- and the counter-passing orbits, the phase space must be divided in disjoint subspaces where the uniqueness of the orbit is recovered. These subspaces are distinguished by an identifier $\sigma$. 


In (\ref{QIave}) the  volume element comes to be 
\begin{equation}
\label{volele}
d^3 x d^3 v = -\frac{w|B|\sqrt{g}}{\psi^\prime |v_\||}d\psi d\theta dw d\lambda d\phi d \gamma=-\frac{wB_{orb}\sqrt{g}}{\psi^\prime }\frac{d\psi d\mathcal{P}_{\phi} dw d\lambda d\phi d \gamma}{|v_\|\partial_\theta \mathcal{P}_{\phi}|},
\end{equation}
with  $\sqrt{g}$ and $\partial_\theta \mathcal{P}_{\phi}$ 
 computed at $\theta=\theta(\psi,\mathcal{P}_\phi,w,\lambda)$, when $|B|=B_{orb}$; $\gamma$ is the gyrophase which is an ignorable coordinate as well as the toroidal angle $\phi$.
In order to demonstrate the equivalence of (\ref{orbavecl}) with (\ref{QIave}), the first equivalence of (\ref{dotpsi}) 
 is used in (\ref{volele}). Finally, the obtained volume element is used into (\ref{QIave}), deducing the equivalence of the QI average with the orbit average (\ref{orbavecl}).
As an application of (\ref{QIave}), here it is the value $\langle \psi \rangle_{orb}$ requested for the biased canonical Maxwellian distribution function (\ref{biasedM}) described in Section (\ref{subsec:problems}): 
\begin{equation}
\label{psiorb}
\langle \psi \rangle_{orb}=\frac{\int_{\psi_{min}}^{\psi_{max}} \psi B_{orb}\sqrt{g}/(\psi^\prime \mathcal{Q} v_\|\partial_\theta \mathcal{P}_{\phi}) \, d\psi}{\int_{\psi_{min}}^{\psi_{max}}  B_{orb}\sqrt{g}/(\psi^\prime \mathcal{Q} v_\|\partial_\theta \mathcal{P}_{\phi}) \, d\psi },
\end{equation}
where $\psi_{min}$ and $\psi_{max}$ are respectively the minimum and the maximum poloidal magnetic flux value reached by the orbit.
Replacing $\psi=\mathcal{P}_\phi - Fv_\|/\omega_c$:
\begin{equation}
\label{psiorb2}
\langle \psi \rangle_{orb}=\mathcal{P}_{\phi}- \frac{m_s\int_{\psi_{min}}^{\psi_{max}} F\sqrt{g}/(\psi^\prime \mathcal{Q} \partial_\theta \mathcal{P}_{\phi}) \, d\psi }{q_s\int_{\psi_{min}}^{\psi_{max}} B_{orb}\sqrt{g}/(\psi^\prime \mathcal{Q} v_\|\partial_\theta \mathcal{P}_{\phi}) \,d\psi },
\end{equation}
where the approximation $\langle \psi \rangle_{orb} \sim \psi_0$ used in \cite{angelino}, can be deduced when $|B| = B_{orb} \sim B_0$.

Another application of the above expression for the time integration is on the computation of the characteristic frequencies, the inverse of the \emph{bounce} time $\tau_b$ of the projected orbits:
\begin{equation}
\label{omegac}
\tau_b \equiv \int_C \, dt=\int \,\delta(\tilde{\mathcal{P}}_{\phi}-\mathcal{P}_{\phi})\delta(\tilde{w}-w)\delta(\tilde{\lambda}-\lambda)\,  \frac{d^3\tilde{x}d^3\tilde{v} }{4\pi^2\mathcal{Q}w} .
\end{equation}
finding out:
\begin{equation}
\label{omegac2}
\fl
\\ \tau_b= -2\int_{\psi_{min}}^{\psi_{max}}\frac{B_{orb}\sqrt{g}}{\psi^\prime \mathcal{Q} |v_\|\partial_\theta \mathcal{P}_{\phi}|}\, d\psi =2 \frac{|q_s|}{wm_s} \int_{r_{min}}^{r_{max}} \, \frac{\sqrt{g}B^3_{orb} dr}{\mathcal{Q} F(2-\lambda B_{orb})|\partial_\theta |B||},
\end{equation}
where the relation $v_\|\partial_\theta \mathcal{P}_{\phi}=-w\tilde{F}B^{-2}[2-\lambda|B|]\partial_\theta |B|$ has been used. In the above integral $\sqrt{g}$, $\mathcal{Q}$ and $\partial_\theta |B|$ have to be computed for $\theta=\theta(\psi,\mathcal{P}_\phi,w,\lambda)$, when $|B|=B_{orb}$. Moreover, $r_{min}(\psi_{min})$ and $r_{max}(\psi_{max})$ are respectively the minimum and the maximum radius (poloidal magnetic flux) value reached by the orbit. 

It is worth noting that the \emph{bounce frequency} and the \emph{transit frequency} are the same frequency $2\pi /\tau_b(\mathcal{P}_\phi,w,\lambda)$ computed respectively for  $\lambda>\lambda_c$ and for $\lambda<\lambda_c$\footnote{The difference between the co-passing and the counter-passing orbits is on the integration range of  the radial coordinate.}. One can show how the critical frequency $2\pi/\tau_b(\mathcal{P}_\phi,w,\lambda_c)$ goes to zero, if $\lambda_c$ is allowed once $\mathcal{P}_\phi$ and $w$ are given. The limiting case of the orbit behavior in close analogy to the pendulum,  analyzed in details by Brizard A. J. et al.\cite{brizard11}, can be recovered, as shown by Chiu S. C. et al.\cite{chiu}.

A quick remark is concerning the \emph{second invariant} $J$, often used as COM instead of $w$, with the property: $\partial_w J =\tau_b$. It is clear from (\ref{omegac}) how it can be expressed in terms of $\mathcal{P}_\phi,w$ and $\lambda$:
\begin{equation}
\label{secondInv}
J =\int_{\tilde{w}<w} \, \delta(\tilde{\mathcal{P}}_{\phi}-\mathcal{P}_{\phi})\delta(\tilde{\lambda}-\lambda)\, \frac{d^3\tilde{x}d^3\tilde{v} }{4\pi^2\mathcal{Q}\tilde{w}}  .
\end{equation}


\subsection{How to select only confined Guiding Centers}
\label{subsec:lost}
The reasoning for considering only confined orbits is usually fairly structured \cite{hsu,egedal}. A simple condition that may be implemented in the codes, as done in \cite{belova}, is analyzed.
For simplicity, it is assumed that $\nabla |B| \neq 0$: for each $r$ (each $\psi$) there exists only one maximum $HFS(r)$ and one minimum $LFS(r)$ of $|B|(r,\theta)$ and these are the only stationary points when $\partial_\theta |B|=0$ (the magnetic flux surface is convex).  In this way it is possible to define an equator as the locus where $|B|=HFS(r)$ on the High Field Side, passing through $|B|=B_0$ at $r=0$, and where $|B|=LFS(r)$ on the Low Field Side. The value of $\theta$ can be fixed to $\theta=0$ in the LFS direction and to $\theta=\pi$ in the opposite (HFS) direction. These rules become evident for the up-down symmetric plasma where the plane of symmetry is the equatorial plane.

As a starting point it is analyzed the range of variability of the following quantities: $\lambda \in [0,1/B_{min}],  |B| \in [B_{min},B_{max}]$ where $B_{max}=\max_{(r,\theta)} |B|$ and $B_{min}=\min_{(r,\theta)} |B|$. Moreover, $\psi \in [\psi_a,0]$  corresponds to $r \in [0,a]$, being $\psi(r)\mid_{r=0}=0$ and $\psi(r)\mid_{r=a}=\psi_a<0$. Using the expression (\ref{Borb}) of $|B|$ in orbit coordinates the condition $B_{min} \leq B_{orb} \leq B_{max}$ must be always verified.
If it happens that $B_{min} \leq B_{orb}(\psi_a,\mathcal{P}_\phi,\lambda,w) \leq B_{max}$ then the orbit can intersect the boundary of the plasma volume, the $r=a$ surface, and the GC will be lost. 
The condition for the confined GC will be the complementary one, when  one of the two situations occurs:
\begin{equation}
\label{Bloss1}
\mathrm{B}=\{ (B_{orb}(\psi_a,\mathcal{P}_\phi,\lambda,w) < B_{min}) \mbox{ or } (B_{orb}(\psi_a,\mathcal{P}_\phi,\lambda,w) > B_{max})\}.
\end{equation}
The critical case when the orbit is tangent to the surface $r=a$ has been considered as if the GC is lost.
It is possible to rewrite the above condition as follows:
\begin{equation}
\label{Bloss2}
\mathrm{B}=\{ (\lambda > \lambda_{bmax}(\mathcal{P}_\phi,\chi)) \mbox{ or } (\lambda < \lambda_{bmin}(\mathcal{P}_\phi,\chi))\},
\end{equation} 
because the maximum and the minimum value of $\lambda$ at the boundary surface $r=a$ are respectively:
\begin{equation}
\label{lbmax}
\lambda_{bmax}=\frac{\chi\tilde{F}_a^2}{[\mathcal{P}_{\phi}-\psi_a]^2B_{min}}\left\{ \left\{1+\frac{2[\mathcal{P}_{\phi}-\psi_a]^2}{\chi\tilde{F}_a^2}\right\}^{1/2}-1\right\}
\end{equation}
and
\begin{equation}
\label{lbmin}
\lambda_{bmin}=\frac{\chi\tilde{F}_a^2}{[\mathcal{P}_{\phi}-\psi_a]^2B_{max}}\left\{ \left\{1+\frac{2[\mathcal{P}_{\phi}-\psi_a]^2}{\chi\tilde{F}_a^2}\right\}^{1/2}-1\right\},
\end{equation}
where $\tilde{F}_a=\tilde{F}(\psi_a)$.
The condition (\ref{Bloss1}) is easily visualized as satisfied in Figures \ref{loss1}(a) and (c), in Figure \ref{loss2}(b) and for the Figures \ref{loss3}(a),(b) and (c).

In the favorable case in which a particular set of $(\mathcal{P}_\phi,w,\lambda)$ selects only a single orbit, the condition (\ref{Bloss1}) or (\ref{Bloss2}) provides that the orbit will be confined. The problem arises when the same set of orbit coordinates represents two orbits.
If these orbits are both confined or both loss (hardly), the above condition will be respectively satisfied, as for  Figures \ref{loss1}(a) and (c), or broken. However, when one of the orbits is confined and the other is not, the condition (\ref{Bloss1}) (or (\ref{Bloss2})) is not anymore sufficient; as it is in the cases of Figure \ref{loss1}(b), Figures \ref{loss2}(a) and (c). Fortunately, it is easy to discriminate which of the two orbits is confined. Indeed, it is possible to have two orbits when the surface $\Lambda$ admits one saddle point at $\lambda_c=\Lambda(r_c,\theta_c)$ and $\lambda<\lambda_c$, when the orbits are passing. 
While the critical $\lambda_c$ divides the orbit in passing or trapped, the critical radius $r_c$ divides the passing orbits in two disjoint families of orbits: for $r<r_c$ the orbits are counter-passing for ions and co-passing for electrons, \emph{viceversa} for $r>r_c$.

 It is obvious that the orbit will be confined if the condition 
\begin{equation}
\label{Bloss3}
\mathrm{A}=\{\exists (r_c,\lambda_c) \mbox{ : } (r <r_c(\mathcal{P}_\phi,\chi)) \mbox{ and } (\lambda<\lambda_c(\mathcal{P}_\phi,\chi))\},
\end{equation}
will occur, regardless (\ref{Bloss1}) or (\ref{Bloss2}).

In summary, the condition for taking into account only confined GC is:
\begin{equation}
\label{dconf}
\delta_{confined}= \cases{ 1, & if  $\mathrm{A} \cup  \mathrm{B}$ \\ 0, &  otherwise.} 
\end{equation}

The criterion (\ref{dconf}) is easy to implement in a gyrokinetic code thanks to a \emph{Metropolis} algorithm which is explained below. Gyrokinetic codes provide an initial value to the GC coordinates taking care of the uncorrelated  initialization between a couple of GCs.  For example it is used a particular set of coordinates:
$\bar{Z}=(r,\theta,\phi,v_\|,\mu)$. Then it is necessary to define the subset of confined GC depending on $\mathcal{P}_\phi.w,\lambda$ and $\sigma$:
\begin{equation}
 \mathrm{Q}_{confined}=\{ \mathcal{P}_\phi.w,\lambda,\sigma \mid \delta_{confined}=1  \}.
\end{equation}
Initialization starts with the assignment of an array of values $\bar{Z}= \bar{Z}_1$. From these values it is possible to calculate $\mathcal{P}_{\phi1}=\mathcal{P}_\phi(\bar{Z_1}), w_1=w(\bar{Z_1}), \lambda_1=\lambda_1(\bar{Z_1})$ and $\sigma_1=\sigma(\bar{Z_1})$. If  $Q_1=(\mathcal{P}_{\phi1},w_1,\lambda_1,\sigma_1) \in   \mathrm{Q}_{confined}$ then the value is taken and the initialization proceeds with the next  $\bar{Z}_2$. Otherwise the set of values is rejected and the code returned to reinitialize $\bar{Z}_1$ with another set of values. Up to load all the GCs.


\section{A new parametric distribution function for gyrokinetic equilibria }
\label{sec:good}

Three guidelines are required to proceed on building the equilibrium distribution function $\mathcal{F}_{eq}$: (a) it behaves as a Boltzmann distribution function, $\mathcal{F}_{eq}\propto \exp -\mathcal{E}/T$; (b) $\mathcal{E}$ is function of QIs; (c) $\mathcal{F}_{eq}$ is mathematically tractable. The first request, although only a guideline,  concerns a property that suggests to consider $\mathcal{F}_{eq}$ something more than a simple fitting model distribution function. Indeed, the Boltzmann behavior commonly expresses a physical property: it is expected when the two-point correlations between particles can be neglected \cite{bouchet}.  The second issue is mandatory once assumed a Boltzmann behavior, because (\ref{1})$: \mathcal{F}_{eq}$ has to depend on QIs to be constant in time. The third request is speculative, but very useful.    

In what follows, it is shown how it is possible to arrive at an expression of the energy  as a function of COMs.
If the distribution of particles is described by a local Maxwellian, many of these will have $|\xi|\sim 1/3$ for the reached isotropy.  The same for the SD distribution function.
When it is considered a minority species from ICRH there will be an anisotropy such that it is easier to find a particle with $|\xi|< 1/3$, or else, when they are considered energetic particles coming from a beam there will be an opposite anisotropy such that $|\xi| \sim 1$ is favored.
The ranges $|\xi|\laeq 1/3$ and $|\xi|\sim 1$ are considered with more care.

Here it is convenient to switch on using particle coordinates, although the symbols used for GC coordinates are preserved. As a starting point it is considered the kinetic energy of the single particle:
\begin{equation}
\label{basic}
\frac{v^2}{2}=\frac{v_\phi^2}{2}+\frac{v^2}{2} \left(1- \frac{v_\phi^2}{v^2}\right)=\frac{v_\phi^2}{2}+\frac{v^2}{2}\cos^2 [\alpha(1+\epsilon)],
\end{equation}
with $\sin \alpha = \xi = v_{||}/v$  and $|\epsilon| < 1$ ( $v_\|\sim v_\phi$ when $|B| \sim B_\phi$ as in tokamaks so that $\epsilon \equiv (\arcsin v_\phi/v)/(\arcsin v_\|/v)-1$ is little enough). When $\xi$ is properly small to allow the expansion $\alpha \sim \xi +\xi^3/6$ then the following term can be expanded:
 \begin{eqnarray*}
\cos^2 [\alpha(1+\epsilon)] & \sim & 1-\alpha^2 (1+\epsilon)^2+\frac{1}{3}\alpha^4 (1+\epsilon)^4 \sim \\
                                                & \sim & 1 - \xi^2 (1+\epsilon)^2 +\frac{\xi^4}{3} \epsilon(2+\epsilon)(1+\epsilon)^2 =\\
                                                & = & 1-\frac{3(1+\epsilon)^2}{4\epsilon(2+\epsilon)} + \frac{\epsilon(2+\epsilon)(1+\epsilon)^2}{3}\left[\xi^2-\frac{3}{2\epsilon(2+\epsilon)}\right]^2 =\\
                                                & = & 1-a_0+ b_0\left(\xi^2+c_0-1\right)^2,
\end{eqnarray*}
where $a_0=3(1+\epsilon)^2/[4\epsilon(2+\epsilon)], b_0=\epsilon(2+\epsilon)(1+\epsilon)^2/3$ and $c_0=1-3/[2\epsilon(2+\epsilon)]$, for convenience.
Then, substituting
 \begin{equation}
\label{v_phi_term}
\frac{v_\phi^2}{2}=\frac{(Rv_\phi)^2}{2R^2}=\frac{(L_Z/q_s-\psi)^2}{2(m_sR/q_s)^2},
\end{equation}
 (\ref{basic}) can be rewritten as follows:
 \begin{equation}
\label{basic2}
\frac{v^2}{2}\sim \frac{(L_Z/q_s-\psi)^2}{2(m_sR/q_s)^2}+\frac{v^2}{2}+\frac{v^2}{2}[-a_0+b_0(\lambda|B|-c_0)^2].
\end{equation}
 How it is evident, this relation follows from the balance between the first and the third term, the sum of which gives  an almost zero contribution. Inserting a new parameter $\kappa > \max (1,a_0)$ then adding and subtracting  $\kappa v^2/2$, it is obtained
 \begin{equation}
\label{basic3}
\frac{v^2}{2}\sim\frac{(L_Z/q_s-\psi)^2}{2(m_sR/q_s)^2}+\frac{v^2}{2}(1-\kappa)+\frac{v^2}{2}[\kappa-a_0+b_0(\lambda|B|-c_0)^2].
\end{equation}
 Finally, putting on the LHS the second term it is possible to rescale the energy, now unbalanced between the $\phi$ component and the one orthogonal to it:
  \begin{equation}
\label{basic4}
\frac{v^2}{2}\sim\frac{(L_Z/q_s-\psi)^2}{2\kappa(m_sR/q_s)^2}+\frac{w(\kappa-a_0)}{\kappa}\left[1+\frac{b_0|B|(\lambda-c_0/|B|)^2}{\kappa-a_0}\right].
\end{equation}
A clear functional dependency on $(w,\lambda,L_Z)$ of the energy $\mathcal{E} \sim m_s v^2/2$ has been found when $\alpha \sim \xi +\xi^3/6$ is a good approximation. The case in which $\xi \sim 1$ or $\alpha\sim \pi/2$ is studied on the following. A slightly different expression respect to (\ref{basic}) is assumed:
\begin{equation}
\label{basic5}
\frac{v^2}{2}=\frac{v_\phi^2}{2}+\frac{v^2}{2}\left \{1 - \sin^2 [\frac{\pi}{2}+(\alpha-\frac{\pi}{2})(1-\epsilon)] \right \},
\end{equation}
with the same condition $|\epsilon|<1$.
Expanding the sine up to the forth order on $(\alpha-\pi/2)$ and preserving terms up to the $(1-\xi^2)^2$ order, the same steps as before are followed, to arrive at
 \begin{equation}
\label{basic6}
\frac{v^2}{2}\sim\frac{(L_Z/q_s-\psi)^2}{2(1+\kappa)(m_sR/q_s)^2}+\frac{w(\kappa-a_1)}{(1+\kappa)}\left[1+\frac{b_1|B|(\lambda-c_1/|B|)^2}{\kappa-a_1}\right],
\end{equation}
with $a_1=3(1-\epsilon)^2/[4\epsilon(2-\epsilon)], b_1=\epsilon(2-\epsilon)(1-\epsilon)^2/3$, $c_1=-3/[2\epsilon(2-\epsilon)]$ and $\kappa > \max (1,a_1)$. A similar result is also obtained for $\xi \sim -1$ or $\alpha \sim -\pi/2$.
The above relations are all expressed in particle coordinates. (\ref{basic4}) and (\ref{basic6}) will change once expressed in GC coordinates. Without entering into the details of the GC transformation, it can be assumed that the given expressions are preserved up to the lowest order if transformed in GC coordinates. Five parameters $(\tilde{\mathcal{P}}_{\phi0},\tilde{\lambda}_0,\tilde{\Delta}_{P_\phi}, \tilde{T}_w, \tilde{\Delta}_\lambda)$ that depend on the GC position, are introduced. $L_Z/q_s$ is substituted with $\mathcal{P}_{\phi}$ and $\mathcal{E}$ is divided by $T$ a reference constant temperature per unitary mass,. Finally, both (\ref{basic4}) and (\ref{basic6}) become (after the GC transformation, in GC coordinates):
\begin{equation}
\label{basic7}
\frac{\mathcal{E}}{T}\sim\frac{(\mathcal{P}_{\phi}-\tilde{\mathcal{P}}_{\phi0})^2}{\tilde{\Delta}_{P_\phi}^2}+\frac{w}{\tilde{T}_w}\left[1\pm \frac{(\lambda-\tilde{\lambda}_0)^2}{\tilde{\Delta}_\lambda^2}\right],
\end{equation}
with the squared parenthesis defined positively, such as  $\tilde{\Delta}_{P_\phi}^2, \tilde{T}_w$ and $ \tilde{\Delta}_\lambda^2$.\\
\begin{figure}[htbp]
\begin{center}

\mbox{
\includegraphics[width=7cm]{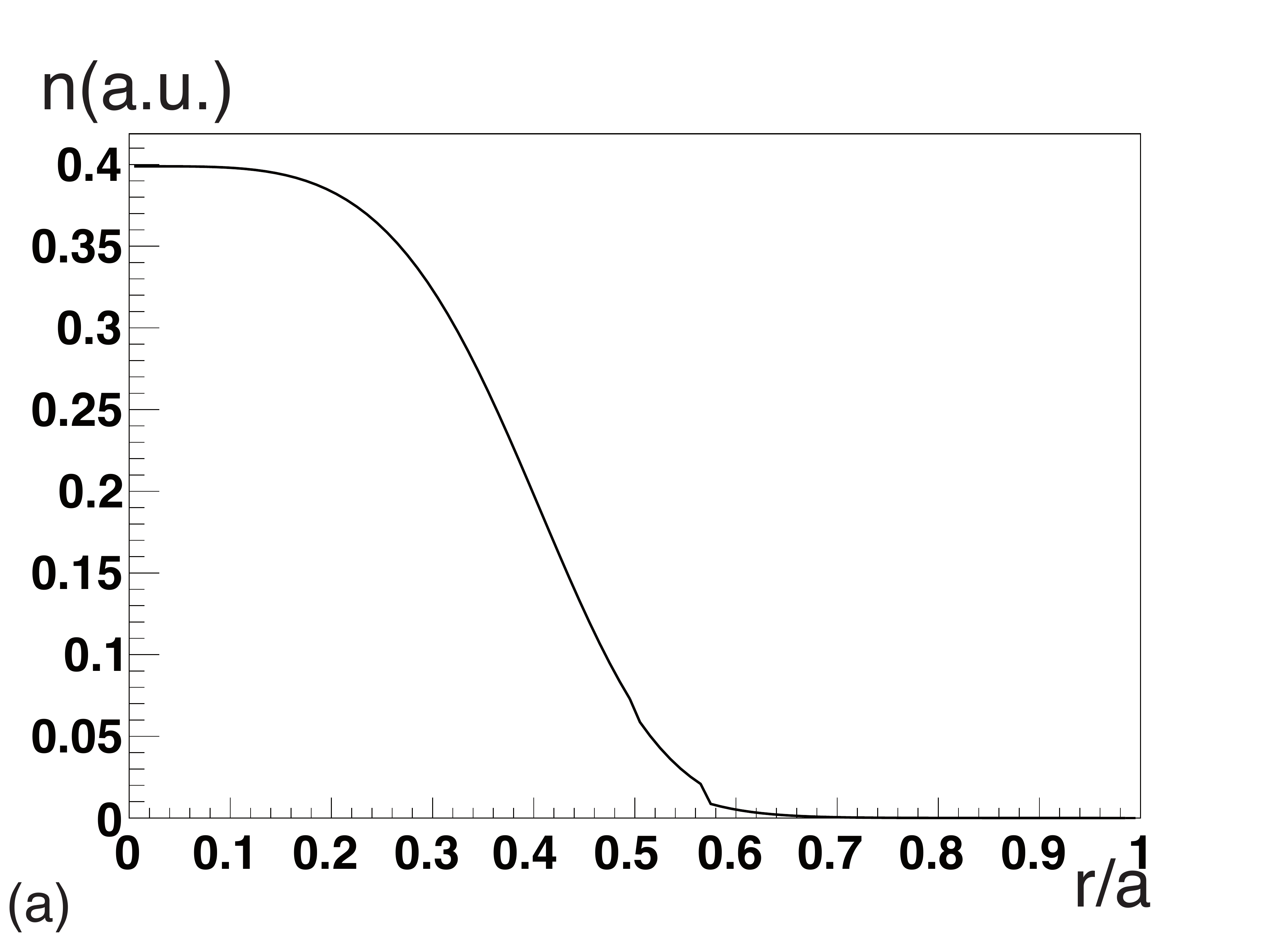}
\includegraphics[width=7cm]{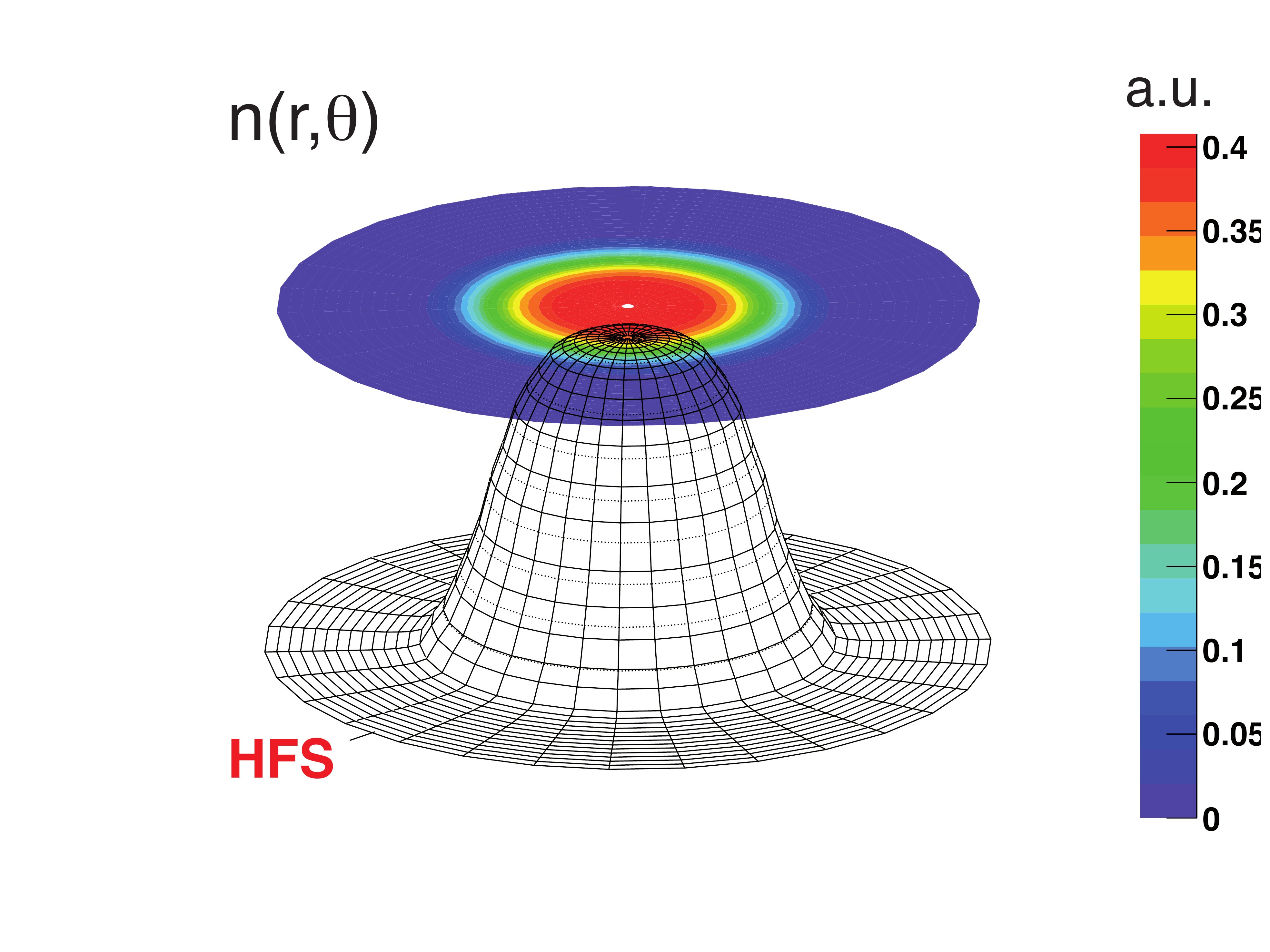}
}
\caption{{\bf (a)} Flux surface averaged density profile $n$ in a.u. versus $r/a$ for the canonical Maxwellian case computed from the dist. func. (\ref{feqlim1}) with $T_w=5.0\times10^{-6}\mbox{ } c^2,\mbox{ } \mathcal{P}_{\phi0}=0.0$ Wb, $\Delta^2_{P_\phi}=0.1$ Wb$^2$. {\bf (b)} Surface polar plot plus contour plot of the density $n$ in a.u. versus $(r,\theta)$ computed from the same dist. func. in (a).}
\label{canoMaxw}
\end{center}
\end{figure}

\begin{figure}[htbp]
\begin{center}
\mbox{
\includegraphics[width=7cm]{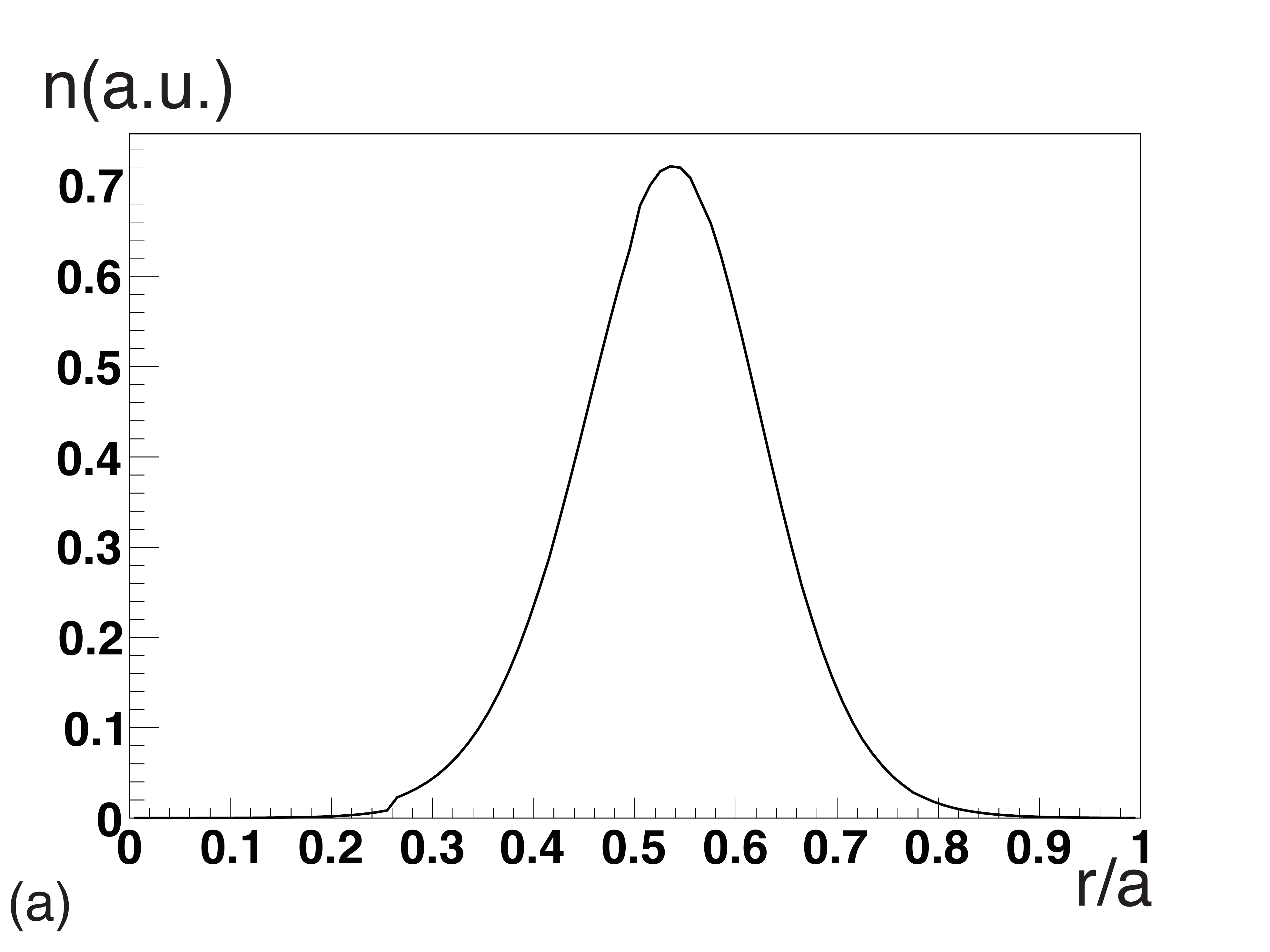}
\includegraphics[width=7cm]{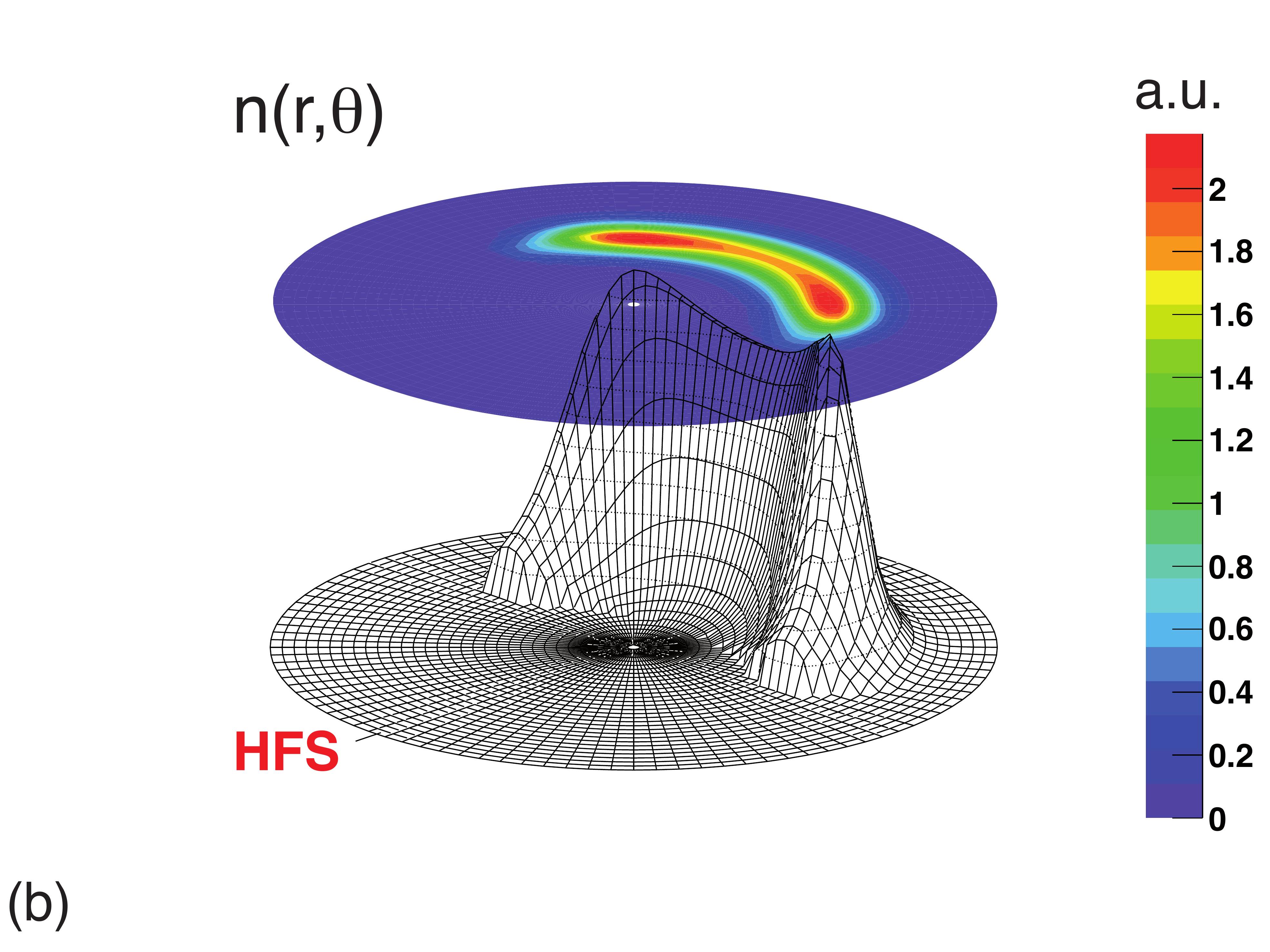}
}
\caption{{\bf (a)} Flux surface averaged density $n$ profile in a.u. versus $r/a$ for the ICRH case computed from the dist. func. (\ref{feq1}) with $\alpha=1.25, \mbox{ } T_w=0.00025\mbox{ } c^2,\mbox{ }  \lambda_0=0.13$ T$^{-1},\mbox{ }  \Delta^2_\lambda=0.00001$ T$^{-2}, \mbox{ } \mathcal{P}_{\phi0}=-1.0$ Wb, $\Delta^2_{P_\phi}=0.05$ Wb$^2$. {\bf (b)} Surface polar plot plus contour plot of the density $n$ in a. u. versus $(r,\theta)$ computed from the same dist. func. in (a).  A lot of GCs are positioned near the $tip$s of the banana orbits, forming two horns.}
\label{denICRH}
\end{center}

\end{figure}

A GC Boltzmann-like equilibrium distribution function can now be written down using (\ref{basic7}) in the exponent. To fulfill the first equilibrium condition (\ref{1}), the functional dependence on the QIs is maintained, whilst the not constant quantities are replaced with constant parameters to be determined afterwards. In this way a parametric equilibrium distribution function is obtained. It leaves to the parameters the task to capture the collective character of how the GCs are distribuited. Indeed, regardless of the plasma operational configuration and of the particle species (fusion products, thermal bulk, energetic particles from ICRH and NNBI), the single particle energy is always given in the form $\mathcal{E}=m_s v^2/2+q_sA_0$. This consideration leads to a unique Boltzmann-like distribution function to be valid in any context. However, ensemble phenomena arise: a collection of these GCs is distributed in a non-uniform fashion; e.g., concentrated in the hot core plasma as for fusion alpha, concentrated off-axis as for ICRH, or having a narrow pitch angle distribution with dominant $v_\|$ as for NNBI. These peculiarities must be taken into account, firstly, separating the perpendicular from the parallel anisotropy component in the kinetic energy (as done above introducing the unbalancing $\kappa$ parameter for the particle kinetic energy expression), then by choosing the appropriate parameters.

As an example, the following distribution function is taken into consideration:
\begin{equation}
\label{feq1}
\fl
\tilde{F}_{eq}=\frac{\mathcal{N}}{\sqrt{2\pi} w^{3/2}}\left( \frac{w}{T_w}\right)^{\alpha}\exp \left[-\left(\frac{\mathcal{P}_{\phi}-\mathcal{P}_{\phi0}}{\Delta_{P_\phi}}\right)^2 \right] \exp \left\{-\frac{w}{T_w}\left[1+ \left(\frac{\lambda-\lambda_0}{\Delta_\lambda}\right)^2\right] \right\},
\end{equation}
where $\mathcal{N}, T_w, \alpha, \mathcal{P}_{\phi0}, \Delta_{P_\phi}, \lambda_0$ and $\Delta_\lambda$ are constant parameters.

\begin{figure}[htbp]
\begin{center}
\mbox{
\includegraphics[width=8cm]{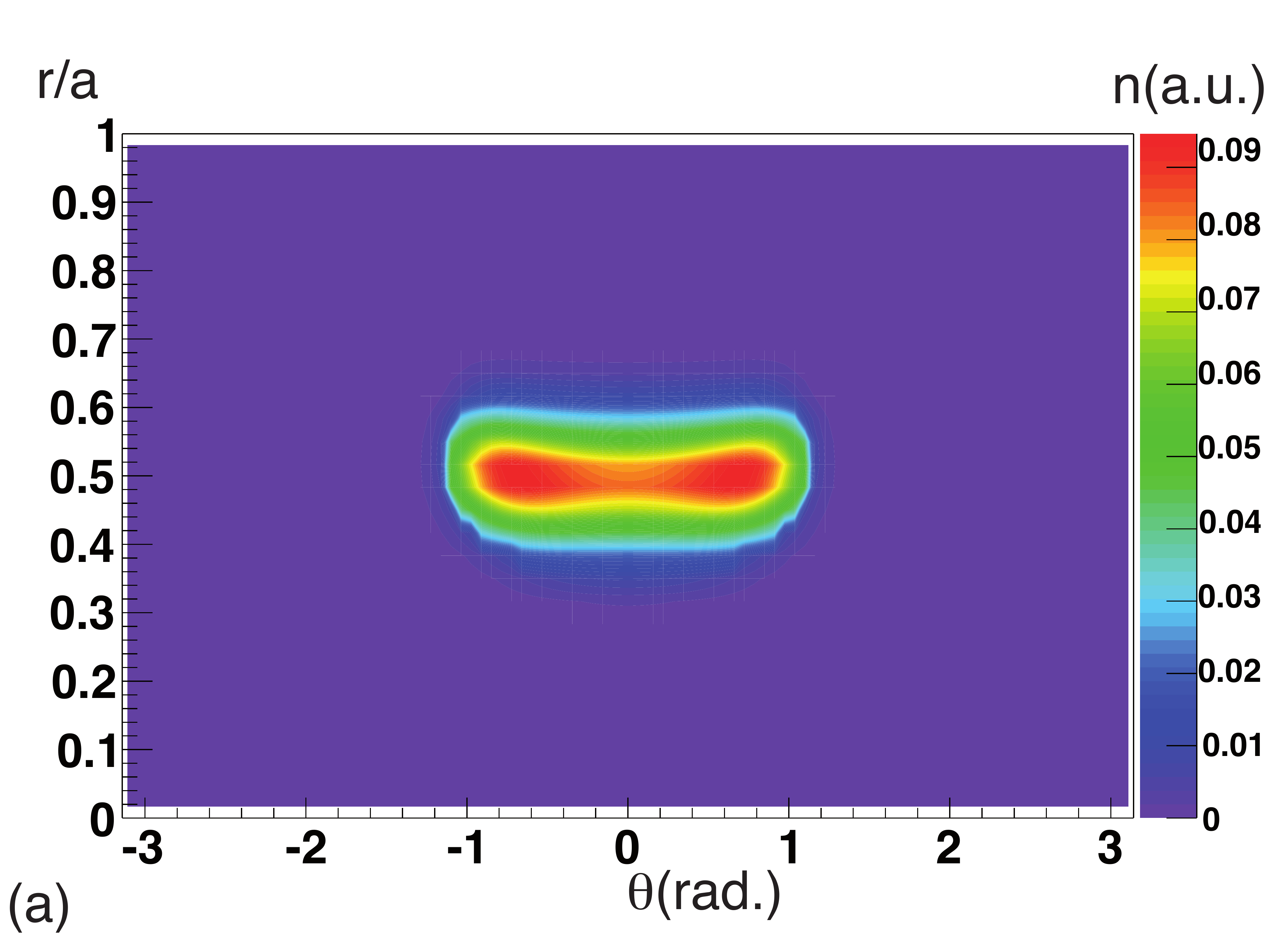}
\includegraphics[width=8cm]{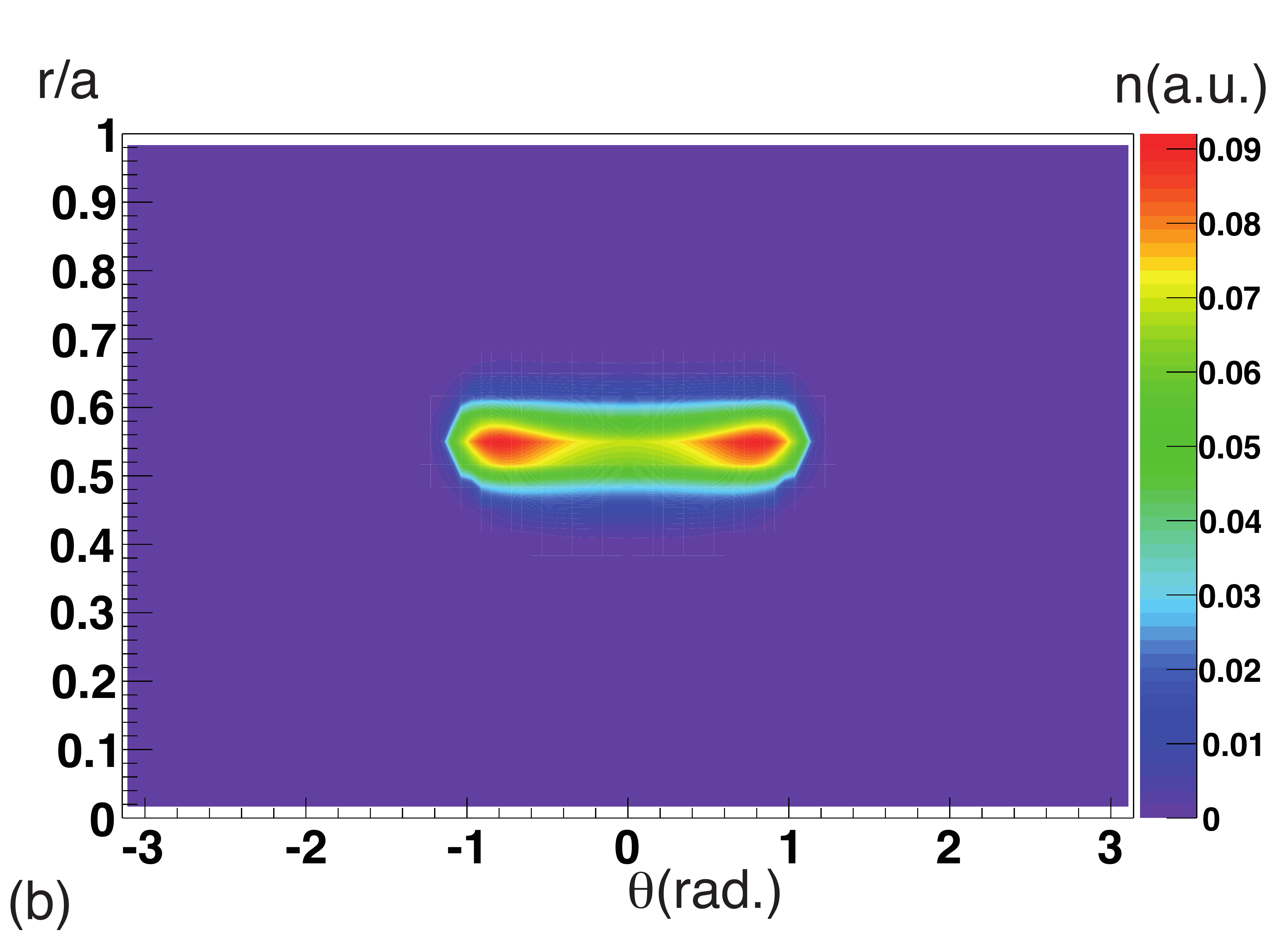}
}
\caption{ Contour plots of the spatial density in $r/a-\theta$ plane for the ICRH case computed from the dist. func. (\ref{feq1}) with: {\bf (a)} $\alpha=1.25, \mbox{ } T_w=0.0002\mbox{ } c^2,\mbox{ }  \lambda_0=0.13$ T$^{-1},\mbox{ }  \Delta^2_\lambda=0.00001$ T$^{-2}, \mbox{ } \mathcal{P}_{\phi0}=-0.85$ Wb, $\Delta^2_{P_\phi}=0.02$ Wb$^2$;  {\bf (b)} $\alpha=1.25, \mbox{ } T_w=0.0002\mbox{ } c^2,\mbox{ }  \lambda_0=0.13$ T$^{-1},\mbox{ }  \Delta^2_\lambda=0.000008$ T$^{-2}, \mbox{ } \mathcal{P}_{\phi0}=-1.0$ Wb, $\Delta^2_{P_\phi}=0.01$ Wb$^2$.}
\label{contICRH}
\end{center}
\end{figure}

The canonical Maxwellian with constant Temperature (and a Gaussian behavior in $\mathcal{P}_\phi$) is obtained for $\Delta_\lambda \to \infty$ and for $\alpha=3/2$, as shown in Figure \ref{canoMaxw}:
\begin{equation}
\label{feqlim1}
\eqalign{
\lim_{\Delta_\lambda \rightarrow \infty} \tilde{F}_{eq}(\alpha=3/2)&=\frac{\mathcal{N}}{\sqrt{2\pi} T_w^{3/2}} \exp \left[-\left(\frac{\mathcal{P}_{\phi}-\mathcal{P}_{\phi0}}{\Delta_{P_\phi}}\right)^2 \right] \exp \left(-\frac{w}{T_w} \right) \\ &=\frac{n(\mathcal{P}_\phi)}{\sqrt{2\pi} T_w^{3/2}} e^{-w/T_w}.}
\end{equation}
A local Maxwellian in the ZOW limit is obtained:
\begin{equation}
\label{feqlim2}
\fl
\lim_{\Delta_\lambda \rightarrow \infty} \tilde{F}_{ZOW}(\alpha=3/2)=\frac{\mathcal{N}}{\sqrt{2\pi} T_w^{3/2}}\exp \left[-\left(\frac{\psi-\mathcal{P}_{\phi0}}{\Delta_{P_\phi}}\right)^2 \right] \exp \left(-\frac{w}{T_w} \right).
\end{equation}
With the case $\Delta_\lambda \to 0, \bar{\mathcal{N}}=\mathcal{N}\Delta_\lambda\neq 0, \alpha=5/4$, (\ref{feq1}) becomes
\begin{equation}
\label{feqlimICRH}
\eqalign{
\lim_{\Delta_\lambda \rightarrow 0} \tilde{F}_{eq}(\alpha=5/4)&=\frac{\bar{\mathcal{N}}}{\sqrt{2} T_w^{3/2}} \left( \frac{T_w}{w}\right)^{3/4} \exp \left[-\left(\frac{\mathcal{P}_\phi-\mathcal{P}_{\phi0}}{\Delta_{P_\phi}}\right)^2 \right] \\ & \exp \left( -\frac{w}{T_w} \right) \delta(\lambda-\lambda_0),}
\end{equation}
which is in accordance with (\ref{singlpit}) in the ZOW limit:
\begin{equation}
\label{feqlim3bis}
\eqalign{
\lim_{\Delta_\lambda \rightarrow 0} \tilde{F}_{ZOW}(\alpha=5/4)&=\frac{\bar{\mathcal{N}}}{\sqrt{2} T_w^{3/2}} \left( \frac{T_w}{w}\right)^{3/4} \exp \left[-\left(\frac{\psi-\mathcal{P}_{\phi0}}{\Delta_{P_\phi}}\right)^2 \right] \\ & \exp \left( -\frac{w}{T_w} \right) \delta(\lambda-\lambda_0).}
\end{equation}
Especially, when the FOW effects are considered, it is observed the expected minority phenomenology in the presence of an ICRH antenna. For example in Figures \ref{denICRH} and \ref{contICRH}, a higher concentration of minority near the banana orbit tips positioned along the resonant value of the magnetic field magnitude $B_{res}=\lambda_0^{-1}$ is shown \cite{kitply}.

Finally, the modified biMaxwellian distribution function (\ref{2M}) can be partially reproduced. Even when FOW effetcs are taken into account such similarities are observed when (\ref{feq1}) is rewritten as 
\begin{equation}
\label{feqlim3bisbis}  
\fl
\tilde{F}_{eq}(\alpha=3/2)=\tilde{n}_h(\mathcal{P}_\phi,\lambda)\frac{m_h^{3/2}}{(2\pi \tilde{T}_\perp(\lambda))^{3/2} } \exp \left\{-m_hw\left[\frac{\lambda B_{res}}{\tilde{T}_\perp(\lambda)}+\frac{1-\lambda B_{res}}{\tilde{T}_\|(\lambda)}\right] \right\}, 
\end{equation}
where the following definition has been adopted:
\begin{eqnarray*}
\tilde{T}_\|(\lambda)&=& \, \frac{m_h\Delta^2_\lambda T_w}{\Delta^2_\lambda+\lambda^2+\lambda^2_0} \\
\tilde{T}_\perp(\lambda)&=& \, \frac{m_h\Delta^2_\lambda T_w}{\Delta^2_\lambda+\lambda^2-\lambda^2_0},
\end{eqnarray*}
and
$$
\fl
\tilde{n}_h(\mathcal{P}_\phi,\lambda)=\frac{2\pi\mathcal{N}\Delta^3_\lambda}{(\Delta^2_\lambda+\lambda^2-\lambda^2_0)^{3/2}}\exp \left[-\left(\frac{\mathcal{P}_\phi-\mathcal{P}_{\phi0}}{\Delta_{P_\phi}}\right)^2 \right].
$$
The similarity with (\ref{2M}) is only when $\lambda<\lambda_0=B^{-1}_{res}$. On the contrary, the following form of the equilibrium distribution function should be considered to obtain (\ref{2M}) when $\lambda>\lambda_0=B^{-1}_{res}$:
\begin{equation}
\label{feq2}
\fl
\hat{F}_{eq}=\frac{\mathcal{N}}{\sqrt{2\pi} w^{3/2}}\left( \frac{w}{T_w}\right)^{\alpha}\exp \left[-\left(\frac{\mathcal{P}_{\phi}-\mathcal{P}_{\phi0}}{\Delta_{P_\phi}}\right)^2 \right] \exp \left\{-\frac{w}{T_w}\left[1- \left(\frac{\lambda-\lambda_0}{\Delta_\lambda}\right)^2\right] \right\},
\end{equation}
with the minus sign in the squared parenthesis of the last exponent. This case can be obtainable from (\ref{basic7}) too.  Even though interesting, in the following the analysis will be focalized on the equilibrium distribution function (\ref{feq1}). 


\subsection{Useful mathematical properties of the proposed distribution function}
\label{subsec:useful}
The equilibrium distribution function (\ref{feq1}) is a totally parametric function particularly indicated for the differential calculus, as can be the application of a differential collisional operator.
 It worths the trouble to emphasize the integral properties of the obtained distribution function mainly used for the velocity momenta computation. 
Furthermore it is important to justify  the choice of the normalization factor proportional to $(w/T_w)^\alpha/\sqrt{2\pi w^3}$) in front of the Boltzmann-like exponential law in (\ref{feq1}).
Going back to the well known variables $(\psi,\theta,w,v_\|)$:
\begin{equation}
\label{change1}
(\mathcal{P}_{\phi}-\mathcal{P}_{\phi0})^2=(\psi-\mathcal{P}_{\phi0})^2+2(\psi-\mathcal{P}_{\phi0})\frac{F}{\omega_c}v_{\|}+\frac{F^2}{\omega_c^2}v_{\|}^2,
\end{equation}
\begin{equation}
\label{change2}
(\lambda-\lambda_0)^2=\frac{(1-\lambda_0|B|)^2}{B^2}-\frac{(1-\lambda_0|B|)v_{\|}^2}{wB^2}+\frac{v_{\|}^4}{4w^2B^2}.
\end{equation}
(\ref{feq1}) can be written as follows:
\begin{eqnarray*}
\label{feq3}
\tilde{F}_{eq}&=&\frac{\mathcal{N}}{\sqrt{2\pi} w^{3/2}}\left( \frac{w}{T_w}\right)^{\alpha}\exp \left[-\left(\frac{\psi-\mathcal{P}_{\phi0}}{\Delta_{P_\phi}}\right)^2\right]\\
 &\exp &\left \{-2(\psi-\mathcal{P}_{\phi0})\frac{F}{\omega_c\Delta_{P_\phi}^2}v_{\|}-\left[\frac{F^2}{\omega_c^2\Delta_{P_\phi}^2}-\frac{(1-\lambda_0|B|)}{T_wB^2\Delta_\lambda^2}\right]v_{\|}^2\right \} \\
 & \exp &\left \{-\frac{[(1-\lambda_0|B|)^2+B^2\Delta_\lambda^2]w}{T_wB^2\Delta_\lambda^2}-\frac{v_{\|}^4}{4wT_wB^2\Delta_\lambda^2}\right \}.
\end{eqnarray*}
In the above expression the $-v_{\|}^4/w$ term is crucial (the sign justifies the choice of the plus instead of the minus sign in (\ref{feq1}) respect to (\ref{feq2})).
When the only $v_\|$ dependence in $\tilde{F}_{eq}$ is considered:
 \begin{equation}
\label{feqv||}
\tilde{F}_{eq}(v_{\|})=\tilde{\mathcal{N}}\exp \left (-\mathcal{J}v_{\|}-\frac{m^2 v_{\|}^2}{2}-\frac{gv_{\|}^4}{4!}\right ),
\end{equation}
with
\begin{equation}
\label{Higgs}
\fl
 \eqalign{ &\tilde{\mathcal{N}}=\frac{\mathcal{N}}{\sqrt{2\pi} w^{3/2}}\left( \frac{w}{T_w}\right)^{\alpha}\exp \left\{-\left(\frac{\psi-\mathcal{P}_{\phi0}}{\Delta_{P_\phi}}\right)^2-\frac{[(1-\lambda_0|B|)^2+B^2\Delta_\lambda^2]w}{T_wB^2\Delta_\lambda^2}\right\}\\
                                  &\frac{m^2}{2}=\left[\frac{F^2}{\omega_c^2\Delta_{P_\phi}^2}-\frac{1-\lambda_0|B|}{T_wB^2\Delta_\lambda^2}\right]\\
                                  &\mathcal{J}=2(\psi-\mathcal{P}_{\phi0})\frac{F}{\omega_c\Delta_{P_\phi}^2}\\
                                  &g=\frac{6}{wT_wB^2\Delta_\lambda^2}. }                               
\end{equation}
The exponent is very similar to the action in the usually called "$\lambda \phi^4$" field theory \cite{zin}, with an interaction term "$\mathcal{J} \cdot \phi$" where $\mathcal{J}$ is the interaction current
(when $m^2<0$ and $\mathcal{J}=0$ the classic double-well potential is recognized).
Thanks to this coincidence, it should be possible to borrow for our use some techniques used in quantum (or condensed matter) field theory. The following formal identity can be useful when $r=(4wT_wB^2\Delta_\lambda^2)^{-1}$ is sufficiently small and $q\neq0$ (or $m^2\neq 0$):
\begin{equation}
\int dx \exp \left(-px-qx^2-rx^4\right)=\sqrt{\frac{2\pi}{q}}\exp\left(-r\partial^4_p\right) \exp\left(\frac{p^2}{4q}\right),
\end{equation}
and again
\begin{equation}
\eqalign{
 &\sqrt{\frac{2\pi}{q}}\exp\left(-r\partial^4_p\right) \exp\left(\frac{p^2}{4q}\right) = \sqrt{\frac{2\pi}{q}}\left(1-r\partial^4_p\right) \exp\left(\frac{p^2}{4q}\right)+\mathcal{O}(r^2) \\ 
&\sim \sqrt{\frac{2\pi}{q}}\left\{1-\frac{3r}{4q^2}\left[ 1+\frac{p^2}{q}+ \frac{p^4}{12q^2}\right]\right\}\exp\left(\frac{p^2}{4q}\right).
}
\end{equation}

 The chosen factor in front of the exponent in (\ref{feq1}) is justified when one explicits the only $w$ dependence. $\tilde{F}_{eq}$ is rewritten as follows:
 \begin{equation}
\label{feqw}
\tilde{F}_{eq}(w)= \frac{\hat{\mathcal{N}}}{\sqrt{2\pi} w^{3/2}}\left( \frac{w}{T_w}\right)^{\alpha} \exp \left(-\frac{aw}{2}-\frac{b}{2w}\right), 
\end{equation}
with
 \begin{equation}
\label{coeff1}
\hat{\mathcal{N}}=\mathcal{N}\exp \left[-\left(\frac{\psi-\mathcal{P}_{\phi0}}{\Delta_{P_\phi}}\right)^2\right]\exp \left (-\mathcal{J}v_{\|}-\frac{m^2 v_{\|}^2}{2} \right),
\end{equation}
and
 \begin{equation}
\label{coeff2}
\eqalign{ & a= 2 \frac{(1-\lambda_0|B|)^2+B^2\Delta_\lambda^2}{T_wB^2\Delta_\lambda^2} \\ &b= \frac{v_{\|}^4}{2T_wB^2\Delta_\lambda^2}.} 
\end{equation}
The function in (\ref{feqw}) is recognized to be the statistical Inverse Gaussian (IG) distribution when $\alpha=0$, otherwise it is proportional to the Generalized Inverse Gaussian (GIG) distribution. The properties of IG and GIG pdfs (\emph{probability density function}s) are well known \cite{life,jorgensen,seshandri}.
For example, the definition of GIG pdf itself is used to show the identity:
\begin{equation}
\label{moments}
\fl
\int_{0}^{\infty} \frac{w^ndw}{\sqrt{2\pi} w^{3/2}}\left( \frac{w}{T_w}\right)^{\alpha} \exp \left(-\frac{aw}{2}-\frac{b}{2w}\right)=\frac{2\mathrm{K}_p(\sqrt{ab})}{\sqrt{2\pi}T_w^\alpha(a/b)^{p/2}},\quad p=n+\alpha-1/2,
\end{equation}
where $\mathrm{K}_p$ is the modified Bessel function of the second kind.

These and other mathematical properties will not be used here. It was important to emphasize the powerful possibilities offered by (\ref{feq1}) when a velocity integration has to be performed.
However, it should be said that the determination of the velocity moments can be obtained in an elementary way only when $\Delta_\lambda$ goes to infinity, otherwise the use of incomplete Bessel functions is required.
\\
\subsection{Energy boundary behavior and confined Guiding Centers condition}
\label{subsec:enbehav}
\begin{figure}[htbp]
\begin{center}
\mbox{
\includegraphics[width=7cm]{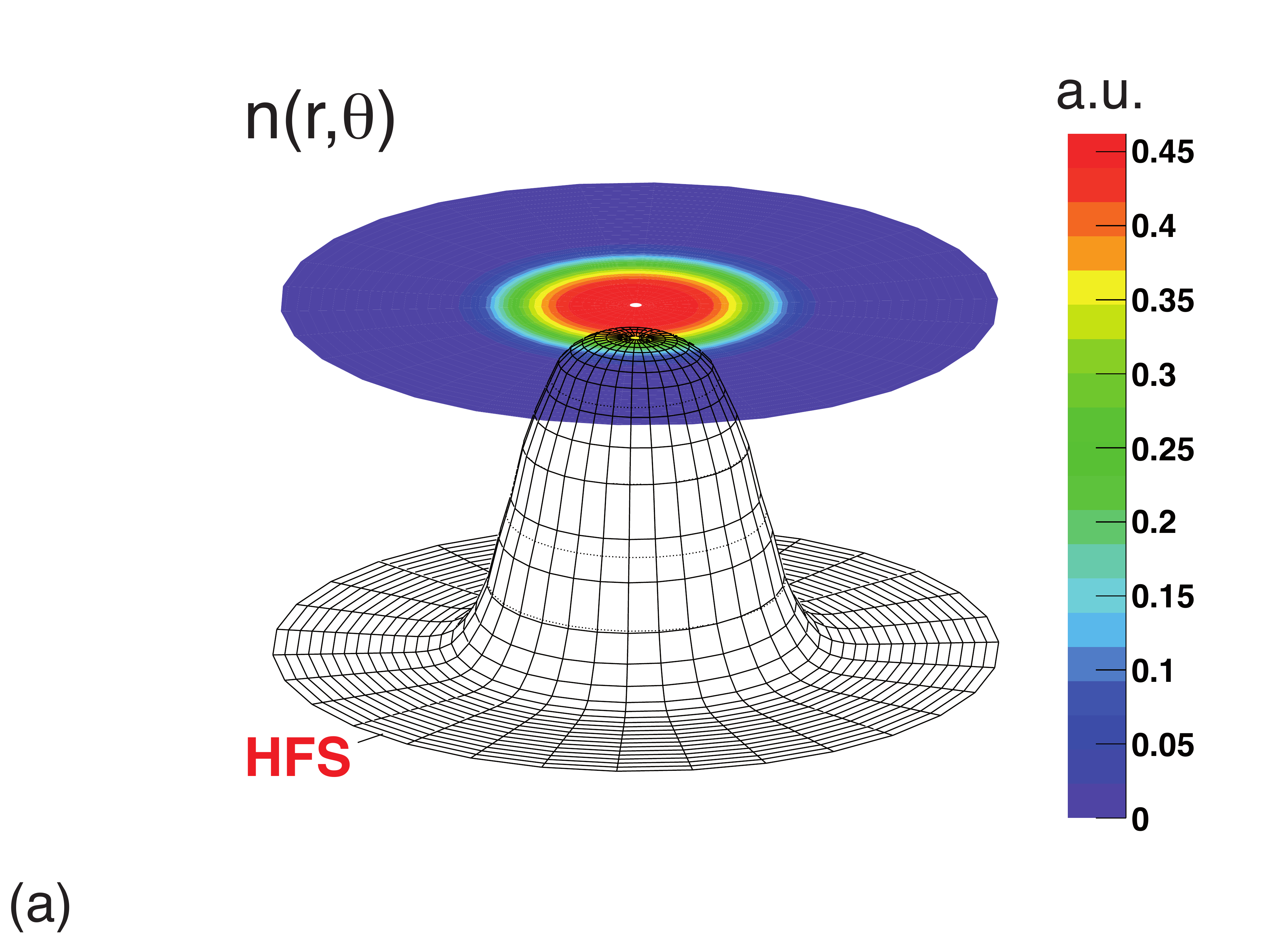}
\includegraphics[width=7cm]{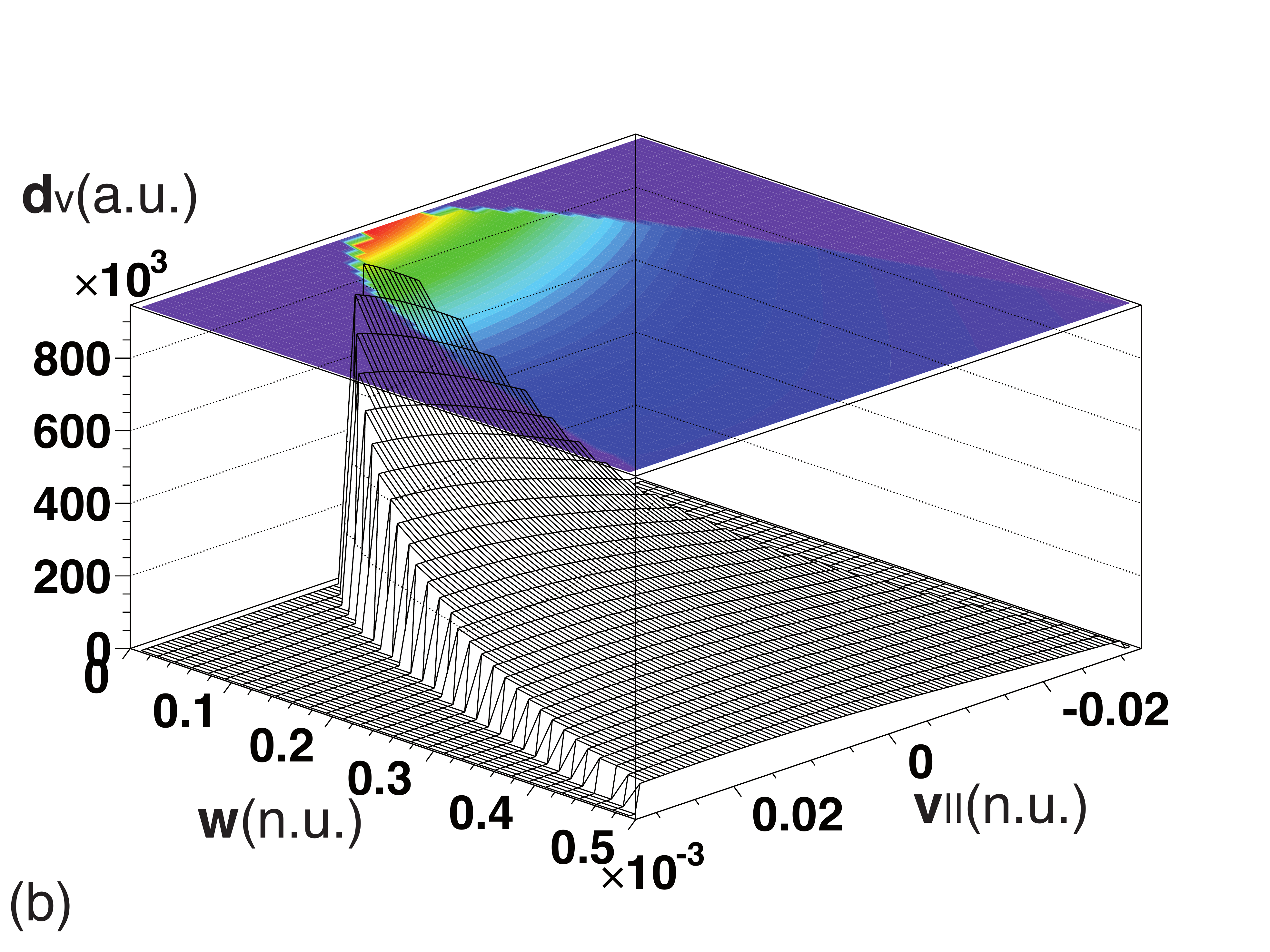}
}
\caption{{\bf (a)} Surface polar plot plus contour plot of the density $n$ versus $(r,\theta)$ in a.u. for the SD case computed from the dist. func. (\ref{feqlim3}) with $\mathcal{P}_{\phi0}=0.0$ Wb, $\Delta^2_{P_\phi}=0.05$ Wb$^2$. {\bf (b)} Velocity density distribution $d_v$ in a.u. versus $(w,v_\|)$ computed from the same dist. func. of (a).  The ridge of the velocity density profile is the standard $w^{-3/2}$ behavior.}
\label{SD}
\end{center}
\end{figure}

The distribution function (\ref{feq1}) gives rise to a logarithmic divergency in the low energy limit (\emph{infrared divergency}) for $\alpha=0$. A simple regularization scheme replacing $w^{3/2}$ with $w^{3/2}+w^{3/2}_0$ is applied to retrieve this interesting case. Looking at (\ref{SD}) and at (\ref{NBI}), the parameter $w_0$ will correspond to the critical energy $w_c$ \cite{sivukhind,stix}. Conversely, the ultraviolet limit ($w\rightarrow \infty$) does not commonly apply because  there is an upper limit on the usable amount of energy. In this way, it is appropriate multiplying the just obtained distribution function for a step function $\mathit{H}(w_1-w)$ (or for some more smoothed sigmoid function). The cut on energy is given at $w_1$. 

At the same time the condition selecting only confined orbits is symbolically represented with the factor  $\delta_{confined}$, defined in (\ref{dconf}).

Finally the following expression for the equilibrium distribution function is obtained:
\begin{equation}
\label{feqdef}
\eqalign{{\mathcal{F}}_{eq}&=\frac{\mathcal{N}}{\sqrt{2\pi} [w^{3/2}+w^{3/2}_0]} \left( \frac{w}{T_w}\right)^{\alpha}\exp \left[-\left(\frac{\mathcal{P}_{\phi}-\mathcal{P}_{\phi0}}{\Delta_{P_\phi}}\right)^2 \right] \\ & \exp \left\{-\frac{w}{T_w}\left[1+\left(\frac{\lambda-\lambda_0}{\Delta_\lambda}\right)^2\right] \right\}\mathrm{H}(w_1-w) \delta_{confined}.
}
\end{equation}

\begin{figure}[htbp]
\begin{center}
\mbox{
\includegraphics[width=7cm]{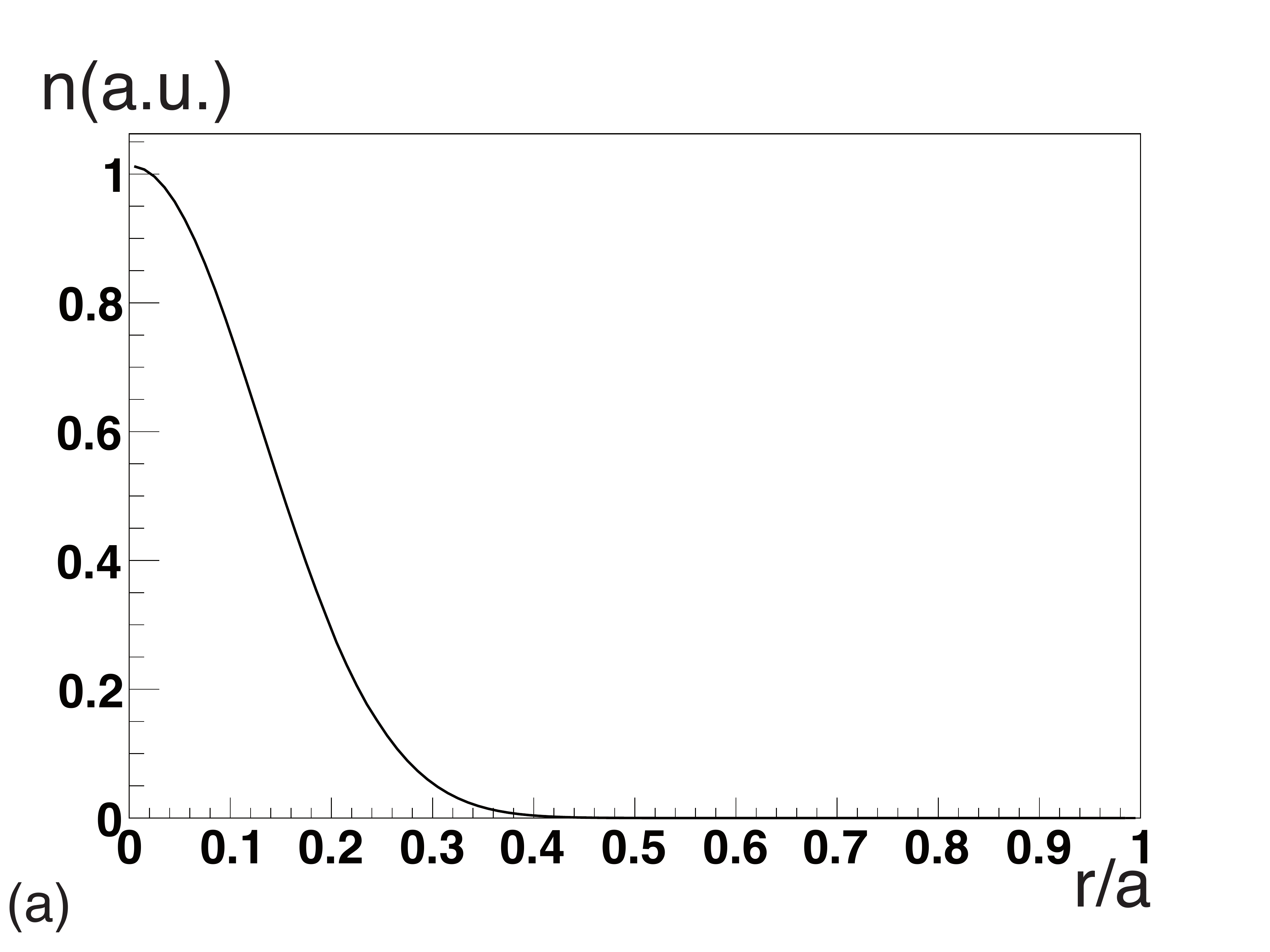}
\includegraphics[width=7cm]{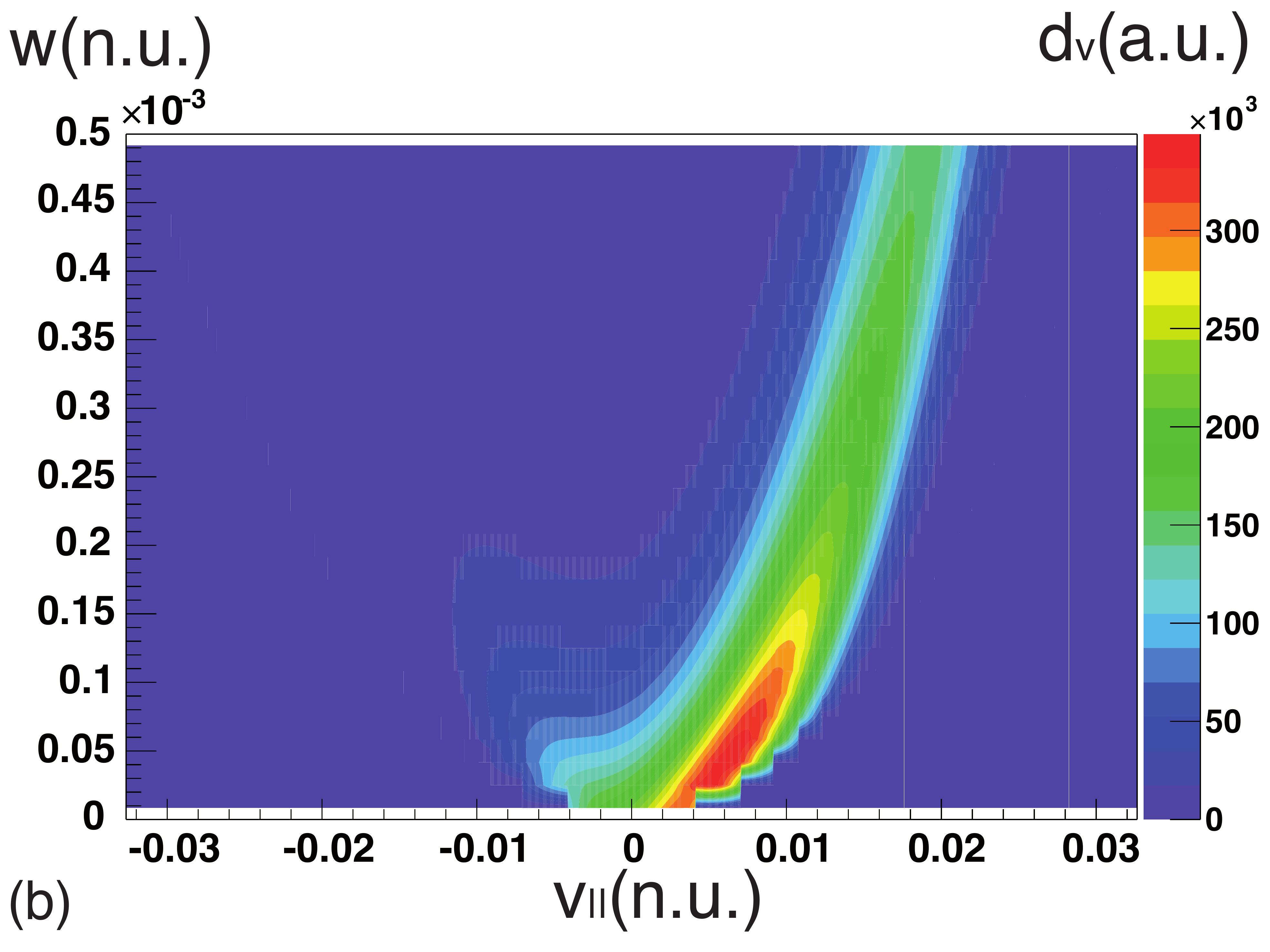}
}
\caption{{\bf (a)} Flux surface averaged density $n$ profile in a.u. versus $r/a$ fot the NNBI case computed from the dist. func. (\ref{feqlim5}) with $\lambda_0=0.08$ T$^{-1},\mbox{ }  T_w \Delta^2_\lambda=5.0 \times 10^{-8}\mbox{ }c^2$ T$^{-2}, \mbox{ } \mathcal{P}_{\phi0}=3.0$ Wb, $\Delta^2_{P_\phi}=0.15$ Wb$^2$.  {\bf (b)}  Contour plot of the velocity density distribution $d_v$ in a.u. versus $(w,v_\|)$ computed from the dist. func. of (a). }
\label{ASD}
\end{center}
\end{figure}

It is possible to find some set of parameters useful to describe a slowing down of the energy. For $T_w \rightarrow \infty$ and for $ \Delta_\lambda \neq 0, \alpha=0$, the SD-like distribution function (\ref{SDdist}) is obtained with a GC anisotropy caused by having a $\mathcal{P}_\phi$ (in place of $\psi$) dependency: 
\begin{equation}
\label{feqlim3}
\fl
\lim_{T_w \rightarrow \infty} \mathcal{F}_{eq}(\alpha=0)=\frac{\mathcal{N}}{\sqrt{2\pi}  [w^{3/2}+w^{3/2}_0]}\exp \left[-\left(\frac{\mathcal{P}_\phi-\mathcal{P}_{\phi0}}{\Delta_{P_\phi}}\right)^2 \right]\mathrm{H}(w_1-w)\delta_{confined}.
\end{equation}
Figure \ref{SD}(a) shows the spatial density in polar coordinates computed from (\ref{feqlim3}). Figure \ref{SD}(b) shows the velocity density distribution in the $(w,v_\|)$ coordinates for the same distribution function. (\ref{feqlim3}) becomes a SD in the ZOW limit:
\begin{equation}
\label{feqlim4}
\fl
\lim_{T_w \rightarrow \infty} \mathcal{F}_{ZOW}(\alpha=0)=\frac{\mathcal{N}}{\sqrt{2\pi}  [w^{3/2}+w^{3/2}_0]}\exp \left[-\left(\frac{\psi-\mathcal{P}_{\phi0}}{\Delta_{P_\phi}}\right)^2 \right]\mathrm{H}(w_1-w).
\end{equation}

Even more possibilities arise when  $T_w \rightarrow \infty$ and $\Delta_\lambda \rightarrow 0$ such as $T_w \Delta_\lambda^2 \neq 0$ is finite. When $\alpha=0$, it is possible to obtain a distribution function that allows the modeling of ions heated by a NNBI:
\begin{equation}
\label{feqlim5}
\eqalign{
\lim_{T_w \rightarrow \infty, \Delta_\lambda \rightarrow 0} \mathcal{F}_{eq}(\alpha=0)&=\frac{\mathcal{N}}{\sqrt{2\pi}  [w^{3/2}+w^{3/2}_0]}\exp \left[-\left(\frac{\mathcal{P}_\phi-\mathcal{P}_{\phi0}}{\Delta_{P_\phi}}\right)^2 \right]\\ & \exp \left[-\frac{w}{T_w}\left(\frac{\lambda-\lambda_0}{\Delta_\lambda}\right)^2 \right]\mathrm{H}(w_1-w) \delta_{confined}.
}
\end{equation}
The "perpendicular anisotropy" term $\exp[-w(\lambda-\lambda_0)^2/(\sqrt{T_w}\Delta_\lambda)^2]$, together with the "parallel anisotropy" term $\exp[-(\mathcal{P}_\phi-\mathcal{P}_{\phi0})^2/\Delta_{P_\phi}^2]$, determines the desired anisotropy in the presence of a NNBI source. The $\lambda$-term determines how many passing orbits are loaded in respect to the unwanted trapped orbits: $\lambda_0$ must be less than $\lambda_c$. The $\mathcal{P}_\phi$-term tunes the balance between co- and counter- passing orbits. Figure \ref{ASD}(b) shows the velocity distribution as function of $(w,v_\|)$ in arbitrary units: the energy follows the behavior $w\sim v_\|^2/2$ which indicates that most of the GCs are passing (in the same plot it is shown an imbalance for the co-passing orbits). Figure \ref{ASD}(a) shows the flux surface average of the density even obtained from (\ref{feqlim5}).
It is worth noting that the anisotropy is mainly due to $\lambda$ in place of $\xi$ with respect to (\ref{NBI}).  A similar choice has already been proposed by \cite{gore} where the Figure 1 therein confirms the result in this article of having a pitch angle width $\tilde{\Delta}_\lambda$ that depends on energy:
$$
\exp \left[-\left(\frac{\lambda-\lambda_0}{\tilde{\Delta}_\lambda}\right)^2 \right]=\exp \left[-\left(\frac{\lambda-\lambda_0}{\sqrt{T_w/w}\Delta_\lambda}\right)^2 \right].
$$


\section{Qualitative description of the equilibrium distribution}
\label{sec:qual}
The factors present in (\ref{feqdef}) are here described in a qualitative way. $\mathcal{F}_{eq}$ is rewritten here for clearness:
\begin{equation}
\label{feqdef2}
\eqalign{
{\mathcal{F}}_{eq}&=\frac{\mathcal{N}}{\sqrt{2\pi} [w^{3/2}+w^{3/2}_0]} \left( \frac{w}{T_w}\right)^{\alpha}\exp \left[-\left(\frac{\mathcal{P}_{\phi}-\mathcal{P}_{\phi0}}{\Delta_{P_\phi}}\right)^2 \right] \exp \left(-\frac{w}{T_w}\right)\\ & \exp \left[ -\frac{w}{T_w}\left(\frac{\lambda-\lambda_0}{\Delta_\lambda}\right)^2 \right]\mathrm{H}(w_1-w) \delta_{confined}.
}
\end{equation}
The following terms can be distinguished.

The Maxwellian term $\exp(-w/T_w)$ is well known, when $\Delta_\lambda \to \infty,\Delta_{P_\phi} \to \infty$ and $w_1 \to \infty$ then $T_w$ becomes the temperature (per unit mass) expressing the decay rate of the energy. For $w_0=0$ and $\alpha=3/2$ the Maxwell distribution function is retrived.

The term $(2\pi)^{-1/2}\mathcal{N}[w^{3/2}+w^{3/2}_0]^{-1}$ indicates the SD of the energy. This behavior is widely known and occurs, for an appropriate range of energies, when modeling the SD distribution function for fusion alpha particles and also for the NBI distribution function for supra-thermal ions (in which cases $w^{3/2}_0 \sim (n_i/n_e)v^3_c/2^{3/2}$ with $v_c$ the critical velocity \cite{sivukhind,stix}). When the important range of energies is much higher then $w_0$, this can be ignored leaving the standard factor $\mathcal{N}/\sqrt{2\pi w^3}$ of the IG pdf. $\mathcal{N}$ is the overall normalization constant which cannot be given explicitly because it is hard to estimate the integration of $\mathcal{F}_{eq}$ on the entire configuration space despite the useful mathematical properties previously shown. 

The term $(w/T_w)^{\alpha}$ is a power law used to take into account mainly the low energy behavior. It is fundamental to evaluate the content of energy that decreases when $\alpha$ increases. When $\alpha=0$ and  $T_w \to \infty,\Delta_{P_\phi} \to \infty$ the SD distribution function (\ref{SDdist}) is retrieved.
When $\alpha=5/4$ and $\Delta_\lambda \to 0$, similarities with (\ref{singlpit}) are found. When $\alpha$ is an integer and $w_0=0$ then it is possible to express the integrals on $w$ with analycal combination of error functions.

The Heaviside step function $\mathrm{H}(w_1-w)$ is used when the particles described are created with a given energy: $w_1$, e.g. $w_1\sim 3.52 $ MeV for the alpha particles. In the presence of a beam, this term  derives  from the monochromatic source approximation ($ \propto \delta(w-w_1)$). When the beam cannot be considered monochromatic, as for source $\propto \exp [-(w-w_1)^2/\Delta_w^2]$, then it would be better using $\mbox{erfc}[(w-w_1)/\Delta_w]$ instead of $\mathit{H}(w_1-w)$.

The term $\exp[-w(\lambda-\lambda_0)^2/(T_w\Delta_\lambda^2)]$ is a completely new term. The  presence of the energy, together with the generalized pitch angle, prevents the factorization of the equilibrium distribution function as $ \propto f(\mathcal{P}_{\phi})g(w)h(\lambda)$. It is worth noticing the  fundamental exception of the single pitch angle case, retrieved as the limit of $\sqrt{w/(T_w\Delta_\lambda^2)} \exp[-w(\lambda-\lambda_0)^2/(T_w\Delta_\lambda^2)]$ for $\sqrt{T_w}\Delta_\lambda/w \to 0$. The relevance of $\lambda$ on classifying the topologies of the orbits has been pointed out above. This term can also counts the number of trapped orbits respect to the passing one. If there is an  interest in studying the behavior of the passing orbits only, then  $\lambda_0 \sim 0$ has to be properly set (deeply passing orbits limit is when $\lambda \to 0$). Moreover, to study only trapped orbits, one has to properly set $\lambda_0 \sim 1/B_{res}$ for a given resonant intensity magnetic field. In this case a large amount of GCs is deposited into the region with $B \sim B_{res}$. For example, this can be the case for simulating the minority ($m_m,q_m$) in the ICRH scheme, for which $\omega_{ICRH}=q_m B_{res}/m_m$.

The $\exp [-(\mathcal{P}_{\phi}-\mathcal{P}_{\phi0})^2/\Delta^2_{P_\phi}]$ term is noteworthy and clear. $\mathcal{P}_{\phi}$ represents the projection of the orbit at given $w$ and $\lambda$ (and $\sigma$). This means orbits are distributed in the most simple way, that is with a Gaussian around a mean value $\mathcal{P}_{\phi0}$. In the SOW case $\mathcal{P}_{\phi} \sim \psi \propto r^2$ in the proximity of the magnetic axis. The tail of this distribution will go rapidly to zero as $\exp -(r/\Delta_r)^4$ when $\mathcal{P}_{\phi0}\sim 0$. On the contrary, when $\mathcal{P}_{\phi0}$ is taken outside the range of the allowed values of $\psi$, the presence of only passing orbits is facilitated (see Figures \ref{loss2} (b) and (c), or \ref{loss3} (b) and (c)). In this case, it will result an imbalance on $v_\|$ due to $\mathcal{J}$ in (\ref{Higgs}), such as in the case of studied population coming from a (N)NBI heating.

The last $\delta_{confined}$ term should be used mainly when there are many orbits with a large width that can be lost.  This term can be implemented numerically thanks to (\ref{dconf}) as described in detail in Section \ref{sec:orb}.

\section{Conclusions and perspetives}
\label{sec:concl}
This work addressed the problem of the equilibrium distribution function  in the gyrokinetic theory. It has been defined an equilibrium distribution function of GCs fulfilling  the following conditions: (\ref{1}) it must depend only on invariants of motion and (\ref{2}) GCs must remain confined for suitably long time. The chosen set of invariants is $(\mathcal{P}_\phi,w,\lambda)$. These invariants have been called Quasi Invariants (QIs) as clarified in Section \ref{sec:concepts}. The Section (\ref{subsec:QI}) emphasizes the connection of the expression of $\mathcal{P}_\phi=\psi+Fv_\|/\omega_c$ with the expression of the drift velocity $v_D$ (\ref{vGC}), thanks to the $\phi$ symmetry. Section  \ref{subsec:problems}  summarizes the way how address the currently studied problem concerning the equilibrium in gyrokinetic simulations in view of emphasizing the alternative approach used in this article.
Some results of the orbit theory have been recoverd in  Section \ref{sec:orb} to introduce the reader to the ($\psi,\mathcal{P}_\phi,w,\lambda$) orbit coordinates used for toroidal symmetric plasma. Some results may be considered more accurate, for example a clear visualization of the orbits and their classification due to the surface $\Lambda$ defined in (\ref{Lambda}). In addition it has been proposed a new way to compute the orbit average (\ref{QIave}) that allows to write the formal expression of $\langle \psi \rangle_{orb}$ in (\ref{psiorb}), as well as the characteristic orbit frequency from the bounce time (\ref{omegac2}). The conditions to discern whether the orbits are confined or loss have been determined ((\ref{Bloss2}), (\ref{Bloss3}) and (\ref{dconf})).
In Section \ref{sec:good}, the equilibrium distribution function has been constructed in parametric form with the following three guidelines:(a) a Boltzmann-like distribution function, $\mathcal{F}_{eq}\propto \exp -\mathcal{E}/T$; (b) $\mathcal{E}$ has to be expressed as function of QIs; (c) $\mathcal{F}_{eq}$ has to be mathematically tractable such as being used in integro-differential calculus. 
$\mathcal{F}_{eq}$ has shown some asymptotic behaviors that are typical of some of the most used distribution functions in the gyrokinetic theory for tokamak plasma namely the local and canonical Maxwellian and the Slowing Down distribution function, respectively (\ref{feqlim2}), (\ref{feqlim1}) and (\ref{feqlim4}). Moreover, the obtained $\mathcal{F}_{eq}$ shows analytic similarities with other distribution functions such as the single pitch angle, the anisotropic Slowing Down and the modified biMaxwellian distribution functions, respectively in (\ref{feqlim3}), (\ref{feqlim5}) and (\ref{feqlim3bisbis}). It is given an explanation for the good  comparison of the behaviors wanted from external sources,  thanks to the possibility of selecting the kind of orbits which are mostly loaded: trapped orbit for the ICRH minority distribution function or passing orbits for the suprathermal ions from NNBI source. This can be deduced from the Figure \ref{ASD}(b) showing the velocity density distribution for the anistopic SD distribution function and indicating the suppression of trapped particles respect to the passing one. On the contrary, the \emph{banana} shape for the minority density in the ICRH case in Figures \ref{denICRH}(b) and \ref{contICRH}, is clearly due to the prevalence of trapped orbits.  The Figures \ref{contICRH} show the density contour plots computed for the ICRH case, and they are surprisingly (because it is machine independent) very similar to the ones reported in \cite{kitply}. 
In Section \ref{subsec:useful}, the mathematical properties of $\mathcal{F}_{eq}$ are partially described. These properties arise from the functional behavior on $w$ and on $v_\|$ that corresponds to the well known cases encountered in statistic analysis and in quantum (or condensed matter) field theory.     
A summary of the various factors that constitute the $\mathcal{F}_{eq}$ has been qualitatively provided in Section \ref{sec:qual}.

$\mathcal{F}_{eq}$ can now be used to fit experimental profiles and it could provide a useful tool for experimental and numerical data analysis.  The proposed model distribution function can be easily implemented in gyrokinetic codes because it is basically an analytical function. In  Section (\ref{subsec:lost}) a method is proposed to implement the condition to avoid loss orbit in the loading subroutine of a gyrokinetic code. This model distribution function has already been used to simulate plasma in the presence of external heating sources, as demonstrated in \cite{cardinali} for the ICRH case relating to FAST \cite{pizzuto} plasma conditions, using the HMGC code \cite{sergio}.

The importance of having a functional fully parametric form has not been shown here. However, the analytical advantages of applying differential operators to it (e.g.  the collision operator) are evident as well as the possibility to perform parametric studies. For example, it would be possible to relate all the  results occurring from a gyrokinetic simulation to the values of the six parameters: $\alpha,T_w,\mathcal{P}_{\phi0},\Delta_{P_\phi},\lambda_0$ and $\Delta_\lambda$.  This is what it is commonly required to carry on the benchmarking of codes.

A final point concerns the modeling of the distribution functions out of the equilibrium. Considering the simple case of the evolution of a plasma population distribution function as transiting between close equilibrium states, this can be easily addressed giving a time dependency to the parameters of $\mathcal{F}_{eq}$. While it may seem premature to consider a possible use of  $\mathcal{F}_{eq}$ in transport models, it seems equally clear that  it can be applied to the integrated simulation of plasma scenarios \cite{itm}. The dialogue between the various codes would be highly optimized if the passage of information occurs through the relevant $\mathcal{F}_{eq}$ parameters. 
 

 \ack{}
 This work was supported by the Euratom Communities under the contract of Association
between EURATOM/ENEA. The author would also like to thank his wife S. D'Antonio for the precious suggestions and V. Fusco, G. Fogaccia, G. Pucella and M.-J. Varamo for comments which helped
to improve the manuscript. Useful discussions with A. Bierwage, A. Biancalani,  G. Vlad,   F. Zonca and S. Briguglio  are also acknowledged. All figures have been realized thanks to the object-oriented framework for data analysis ROOT \cite{root}.

\appendix
\section{Magnetic flux and Shafranov coordinate system}
In plasma theory and modeling a very useful and often adopted representation of the magnetic field is the magnetic flux representation.

The Shafranov coordinates are particularly used in the context of  plasmas with circular poloidal flux section geometry. Once the \emph{Grad-Shafranov} equilibrium equation \cite{grad2,shafranov2} is imposed to the system, this coordinate system leads to the \emph{s-}$\alpha$ model \cite{salpha}. 

In this appendix the magnetic field is briefly described in terms of \emph{flux coordinates} before moving to Shafranov coordinates. Moreover, it is shown the correspondence between the two aforementioned representations. 

\subsection{Magnetic field flux coordinates representation}
 In flux coordinates $B$ is simply written as follows:
\begin{equation}
\label{Bflux}
 B=\nabla \psi \times \nabla \phi +\frac{1}{2\pi} \nabla \phi_t \times \nabla \vartheta,
 \end{equation}
where $\phi$ is the toroidal coordinate, $\vartheta$ is the poloidal angle in flux coordinate, $- 2 \pi \psi=\phi_p$ is the poloidal magnetic flux and $\phi_t$ is the toroidal magnetic flux. Thanks to $\nabla \cdot B=0$ and the \emph{Gauss} theorem, the fluxes can be expressed as:
\begin{equation}
\label{rel1a}
\phi_p=\frac{1}{2\pi}\int_{\Omega(\psi)}B \cdot \nabla \vartheta \, d^3x 
\end{equation}
\begin{equation}
\label{rel1b}
 \phi_t=\frac{1}{2\pi}\int_{\Omega(\psi)}B \cdot \nabla \phi \, d^3x ,
 \end{equation}
 where $\Omega(\psi)$ is the plasma volume enclosed into the magnetic flux surface $\psi$.
 The toroidal coordinate system $(r,\vartheta,\phi)$, where the radial coordinate $r$ labels the magnetic flux surface $r=r(\psi)$, is known as the flux coordinate system if   
\begin{equation}
\label{rel1c}
 d \vartheta= d \phi /q(r)  
\end{equation}
along the magnetic field line. (\ref{rel1c})  is known as  the \emph{straight-line} property of the magnetic field line. 
In flux coordinates the volume element $d^3x$ becomes $d^3 x= \sqrt{g_{flux}} dr d\vartheta d\phi$ where  the \emph{Jacobian} is 
\begin{equation}
\label{rel2}
\sqrt{g_{flux}}\equiv(\nabla r \times \nabla \vartheta \cdot \nabla \phi)^{-1}=-\psi^\prime/B\cdot \nabla \vartheta,
\end{equation}
where the \emph{prime} indicates the radial derivative.

In (\ref{rel1c}) $q(r)$ is the \emph{safety factor} which can also be expressed as:
\begin{equation}
\label{safef}
q(r)=\frac{B\cdot \nabla \phi}{B\cdot \nabla \vartheta}=\frac{d\phi_t}{d\phi_p}=-\frac{1}{2\pi}\frac{d\phi_t}{d\psi}.
\end{equation}
From (\ref{safef}) the representation (\ref{Bflux}) is rewritten in the \emph{Clebsh representation}:
\begin{equation}
\label{clebsh}
B=\nabla \psi \times \nabla (\phi-q \vartheta).
\end{equation}
 The general form of the vector potential $A$ is deduced from (\ref{clebsh}). Indeed, $B=\nabla \psi \times \nabla \phi+ \nabla q \vartheta \times \nabla \psi=\nabla \times (\psi \nabla \phi + q \vartheta \nabla \psi)$ and the vector potential $A$ is written in the Clebsh parametrization \cite{clebsh}:
\begin{equation}
A=\psi\nabla\phi+q \psi^\prime \vartheta\nabla r+ \nabla g,
\end{equation}
where $g$ is a \emph{gauge} function\footnote{It is worth noticing that  in this representation $A$ is a multivalued function. This is not a problem because it is consistent with the \emph{gauge} invariance of $B=\nabla \times A$, because $q\psi^\prime (\vartheta+2\pi)\nabla r = q\psi^\prime \vartheta \nabla r - \nabla \phi_t$.}. When the gauge $\partial_\phi g=0$ is chosen, then the toroidal component $A_\phi=A\cdot e_\phi$ is
\begin{equation}
\label{Aphi}
A_\phi=\psi/R,
\end{equation}
because the unit toroidal vector is $e_\phi=R\nabla \phi$.

In section \ref{subsec:QI} another representation of the magnetic field is used. The $B$ toroidal component is written as:
\begin{equation}
\frac{1}{2\pi} \nabla \phi_t \times \nabla \vartheta=F\nabla \phi,
\end{equation}
where $\partial_\phi F=0$ for axisymmetric systems and   $\partial_\vartheta F=0$\footnote{This result can be obtained for a general poloidal angle $\theta$ different from $\vartheta$, following the same steps of the sketched demonstration.} when the condition $\nabla r \cdot J=0$ is imposed on the plasma density current $J=\nabla \times B/ (4 \pi)$. Indeed,
\begin{equation}
\label{Jr}
4 \pi \nabla r \cdot J= \nabla \cdot B \times \nabla r=\nabla F \cdot \nabla \phi \times \nabla r=\partial_\vartheta F/\sqrt{g_{flux}}=0,
\end{equation}
because of the following equivalences:
\begin{equation}
\nabla \cdot (\nabla \psi \times \nabla \phi) \times \nabla r = \nabla \cdot (\psi^\prime |\nabla r|^2 \nabla \phi)= \nabla (\psi^\prime |\nabla r|^2) \cdot \nabla \phi= 0.
\end{equation}
(\ref{Bflux}) is now rewritten as
\begin{equation}
\label{B0}
B=\nabla \psi \times \nabla \phi + F(\psi)\nabla \phi.
\end{equation}

The flux surface average of some function $f(x)$ is defined as
\begin{equation}
\label{fluxave1}
\left \langle f  \right \rangle(r) \equiv \int d^3\tilde{x} f(\tilde{x}) \delta(r-\tilde{r})  \mbox{\Large{/}}   \int d^3\tilde{x}  \delta(r-\tilde{r}).
\end{equation}
In flux coordinates, (\ref{rel2}) allows to write (\ref{fluxave1}) as follows:
 \begin{equation}
 \label{fluxave2}
 \fl
\left \langle f  \right \rangle(r) =\oint f\sqrt{g_{flux}}\, d\vartheta d \phi  \mbox{\Large{/}}\oint \sqrt{g_{flux}}\, d\vartheta d\phi=\oint f\, \frac{d\vartheta d \phi}{B\cdot \nabla \vartheta}  \mbox{\Large{/}} \oint \, \frac{d\vartheta d\phi}{B\cdot \nabla \vartheta} .
\end{equation}

\subsection{Shafranov coordinates}
Shafranov coordinates are useful for axisymmetric toroidal equilibrium geometry characterized by nested flux surfaces with circular cross sections. The difference with the \emph{standard model} \cite{balescu} is on a relative shift of the centers of the circles corresponding to different flux surfaces: the \emph{Shafranov shift} $\Delta(r)$.
\begin{figure}[htbp]
\begin{center}
\includegraphics[width=15cm]{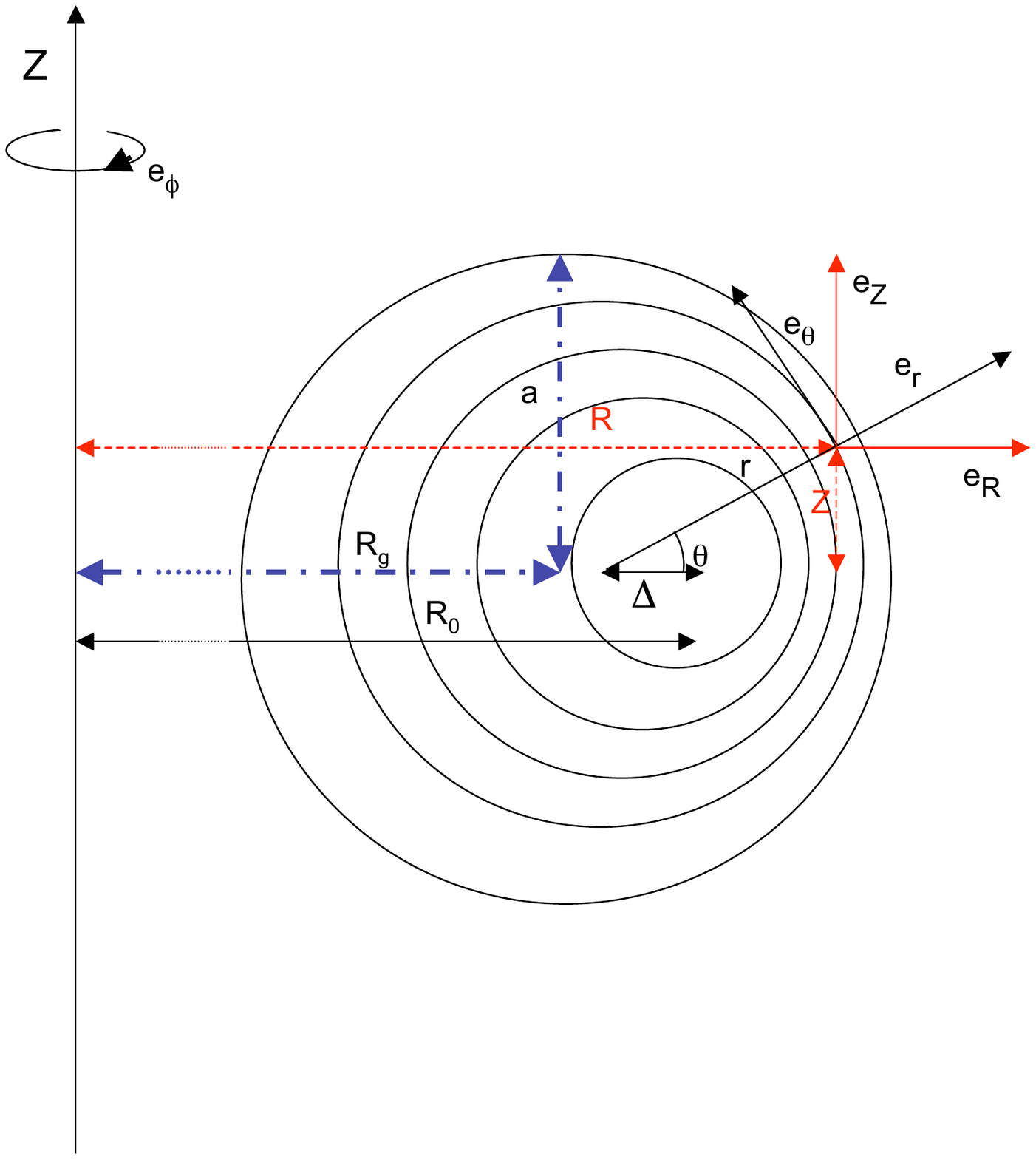}
\caption{Displaced circular magnetic flux surfaces of Shafranov geometry.}
\label{default}
\end{center}
\end{figure}
The map between the cylindrical coordinates $(R,\phi,Z)$ and the Shafranov coordinates $(r,\theta,\phi)$ is the following:
\begin{equation}
\label{syscoshaf}
\left \{\eqalign{ R=R_0-\Delta(r)+r \cos \theta \\
Z= r \sin \theta \\ \phi=\phi,}
\right.
\end{equation}
where $R_0$ is the major radius of the magnetic axis and $\Delta(r)$ is normalized to give $\Delta(0)=0$.  The center $R_g$ of the outermost magnetic flux surface when $r=a$ is obtained if $\Delta(a)=R_0-R_g$. 

This geometry is qualitatively depicted in Figure \ref{default} where the directions of the unit orthogonal vectors $e_r,e_\theta$ and $e_\phi$ respect to the unit vectors $e_R=\nabla R$ and $e_Z=\nabla Z$, are shown.
Once the $\nabla$ operator is applied to (\ref{syscoshaf}), the following relations are obtained:
 \begin{equation}
 \label{versor}
\eqalign{ e_r \equiv e_R \cos \theta + e_Z \sin \theta =(1- \Delta' \cos \theta) \nabla r\\
 e_\theta \equiv -e_R \sin \theta + e_Z \cos \theta=\Delta' \sin \theta \nabla r +r \nabla \theta.}
\end{equation}

Equations (\ref{versor}) can now be reversed requiring  $\Delta^\prime<1$:
\begin{equation}
\label{a15}
\eqalign{\nabla r = (1-\Delta' \cos \theta)^{-1} e_r\\
r \nabla \theta=+e_\theta-\Delta' \sin \theta(1-\Delta'\cos \theta)^{-1}e_r.
 }
\end{equation}
From (\ref{a15}) and using the orthogonality property of the left handed basis $e_r,e_\theta, e_\phi$, it is computed the Jacobian:
\begin{equation}
\label{jacoshaf}
\sqrt{g_{shaf}}\equiv (\nabla r \times \nabla \theta \cdot \nabla \phi)^{-1}= (1-\Delta' \cos \theta)rR.
\end{equation}

Once obtained the Jacobian, the expression for computing the flux surface average (\ref{fluxave1}) of a generic function $f=f(r,\theta)$ is straightforward obtained:
\begin{equation}
\label{fluxave3}
\fl
\left \langle f \right \rangle (r) =\oint f\sqrt{g_{shaf}}\, d\theta\mbox{\Large{/}}\oint \sqrt{g_{shaf}}\, d\theta=\frac{1}{2\pi}\oint f \frac{(1-\Delta' \cos \theta)rR}{( R_0 -\Delta)r- r^2\Delta'/2}\, d\theta,
\end{equation}
because of the following integration:
\begin{equation}
\oint (1-\Delta' \cos \theta)rR\, d\theta=2 \pi \left[( R_0 -\Delta)r- r^2\Delta'/2\right].
\end{equation}

From (\ref{B0}),  the magnetic field is expressed as:
\begin{equation}
\label{B1}
B=\frac{-\psi^\prime}{R(1-\Delta' \cos \theta)} e_\theta+ \frac{F}{R} e_\phi.
\end{equation}
The correspondence with the flux representation is obtained computing the flux poloidal angle $\vartheta$ from the \emph{Shafranov} poloidal angle $\theta$. The relation $\vartheta=\vartheta(\theta,r)$ is obtained from the integration of  $d \vartheta=\partial_\theta \vartheta d \theta+ \partial_r \vartheta d r$ on a constant flux, or constant $r$, path. 

Equations (\ref{B0}) and (\ref{jacoshaf}) give
 \begin{equation}
 \label{partheta}
 \fl
 \partial_\theta \vartheta =\frac {B\cdot \nabla \vartheta}{B\cdot \nabla \theta}=-\frac {B\cdot \nabla \vartheta}{\psi^\prime}\sqrt{g_{shaf}}=-\frac{B\cdot \nabla \phi}{q\psi^\prime}\sqrt{g_{shaf}}=-\frac{rF(1-\Delta' \cos \theta)}{q\psi^\prime R}
 \end{equation}
 and 
 \begin{equation}
\label{vartheta}
\vartheta=\int^{\theta} \partial_{\theta}\vartheta \, d\tilde{\theta} =-\frac{rF}{q\psi^\prime(R_0-\Delta)}\int^{\theta} \frac{1-\Delta' \cos \tilde{\theta}}{1 + r\cos \tilde{\theta} /(R_0-\Delta)}\, d\tilde{\theta}.
\end{equation}
From (\ref{vartheta}) and the $2\pi$ periodicity of the poloidal angles, $\psi^\prime$ can be expressed as 
\begin{equation}
 \label{relpsi'1}
 \psi^\prime=-\frac{rF}{q(R_0-\Delta)}\mathcal{I}(r),
 \end{equation}
where 
\begin{equation}
\label{ident1}
 \eqalign{\mathcal{I}(r) \equiv \frac{1}{2\pi}\oint \frac{1-\Delta' \cos \tilde{\theta}}{1 + r\cos \tilde{\theta} /(R_0-\Delta)}\, d\tilde{\theta}= \\ 
=1+\frac{1}{2\pi}\sum_{k=1}\left( \frac{r}{R_0-\Delta} \right)^{2k-1}\left( \frac{r}{R_0-\Delta}+\Delta^\prime \right)\frac{(2k-1)!!}{2k!!}.}
\end{equation}

When the inverse aspect ratio $\varepsilon=a/R_g$ is little enough and $\Delta^\prime=\mathcal{O}(\varepsilon)$,  the relations (\ref{fluxave3}), (\ref{vartheta}) and (\ref{relpsi'1}) are truncated and respectively approximated by:
 \begin{equation}
\label{fluxave}
\left \langle f \right \rangle (r)=\frac{1}{2\pi}\oint \, d\theta f \{1+[r/(R_0-\Delta)-\Delta'] \cos \theta+\mathcal{O}(\epsilon^2)],
\end{equation}
 \begin{equation}
 \label{varthetatheta}
\vartheta=\theta-[r/(R_0-\Delta)+\Delta']\sin\theta+\mathcal{O}(\epsilon^2),
\end{equation}
\begin{equation}
 \label{relpsi'la}
 \psi^\prime=-\frac{rF}{q(R_0-\Delta)}+\frac{F}{q}\mathcal{O}(\epsilon^3),
 \end{equation}
where it is assumed $\vartheta(\theta=0,r)=0$.

 Moreover, the magnetic field is written as:
\begin{equation}
\label{B2}
B=\frac{rF[1+\mathcal{O}(\epsilon^2)]}{qR(R_0-\Delta)(1-\Delta' \cos \theta)} e_\theta+ \frac{F}{R} e_\phi.
\end{equation}
from which it follows the magnitude used in (\ref{Bshaf}):
\begin{equation}
\label{Bmagnitude}
 |B|=\frac{F}{R}\left[ 1+\frac{r^2}{2q^2(R_0-\Delta)^2}+\mathcal{O}(\epsilon^3)\right]. 
\end{equation}

Sometimes it is preferred the usage of geometrical coordinates when the poloidal sections of  the flux surfaces are circular as the examined case. The transformation map can be obtained by (\ref{syscoshaf}) with the following relations:
\begin{equation}
\label{syscogeom}
\left \{\eqalign{ R=R_g+r_g \cos \theta_g \\
Z= r_g \sin \theta_g \\ \phi=\phi.}
\right.
\end{equation}
 The geometrical coordinates $(r_g,\theta_g)$ can be expressed in terms of  the Shafranov coordinates $(r,\theta)$ by the following reletions:
\begin{equation}
\label{map}
\left \{\eqalign{ r_g^2=r^2+2r\Delta_g(r) \cos \theta + \Delta_g^2(r) \\
\cot \theta_g =\cot \theta +\Delta_g(r)/(r\sin \theta),}
\right.
\end{equation}
with $\Delta_g(r)=R_0-R_g-\Delta(r)$.
When $\Delta_g(r)/r\ll 1$ it is possible to approximate $r\sim r_g -\Delta_g(r_g) \cos \theta_g$. In this case it is straightforward to re-write the relations   (\ref{fluxave}, \ref{relpsi'la}, \ref{varthetatheta}, \ref{B2}, \ref{Bmagnitude}) as function of $(r_g,\theta_g)$. As for example the poloidal flux expressed in geometrical coordinates $\psi(r)=\psi_g(r_g,\theta_g)$ becomes:
\begin{equation}
\label{psigeom}
\eqalign{ \psi(r)=\int^r \psi^\prime \,d \tilde{r} \sim \int^{r_g-\Delta_g \cos \theta_g} \psi^\prime \,d r \sim \\ 
\sim \int^{r_g} \psi^\prime \,d r -\Delta_g \cos \theta_g \psi^\prime(r_g)= \psi(r_g) -\Delta_g \psi^\prime(r_g) \cos \theta_g,}
\end{equation}
 so that the \emph{Fourier} representation of the equilibrium flux potential $\psi_g \sim \psi(r_g) -\Delta_g \psi^\prime(r_g) \cos \theta_g$ in geometrical coordinates involves only the first harmonic. It is worth noticing that (\ref{psigeom}) depends on the geometry, not on equilibrium constrains as can be  the application of the \emph{Grad-Shafranov} equation.  Moreover, (\ref{psigeom}) fails on describing a system near to the magnetic axis where $\Delta_g(r)/r\ll 1$ does not occur because $\Delta_g(0)=R_0-R_g \ne0$ is assumed. 
  
  The reader interested in the generalization of the Shafranov coordinates for describing axisymmetric  plasmas subject to the \emph{Grad-Shafranov} equation in shaped poloidal magnetic flux surface sections, can consult for example \cite{deblank2}.

\end{document}